\documentclass[useAMS,usenatbib]{mn2e}
\usepackage{graphicx,amssymb, txfonts}

\def\pb{Pa$\beta$}
\def\br{Br$\gamma$}
\def\ha{H$\alpha$}
\def\hb{H$\beta$}
\def\feii{[Fe\,{\sc ii}]}
\def\pii{[P\,{\sc ii}]}
\def\oiii{[O\,{\sc iii}]}

\def\h2{H$_2$}
\def\p1{Paper~I}

\def\kms {$\rm km\,s^{-1}$}

\title[Feeding and feedback in Mrk\,1157]{Feeding and feedback in the active nucleus of Mrk 1157 probed with Gemini NIFS}

\author[Riffel \& Storchi-Bergmann]{Rogemar. A. Riffel$^{1}$\thanks{E-mail:
rogemar@ufsm.br} and Thaisa Storchi-Bergmann$^{2}$\\
$^{1}$ Universidade Federal de Santa Maria, Departamento de F\'\i sica, Centro de Ci\^encias Naturais e Exatas, 
97105-900, Santa Maria, RS, Brazil\\
$^{2}$Universidade Federal do Rio Grande do Sul, Instituto de F\'\i sica, CP 15051, Porto Alegre 91501-970, RS, Brazil
}
\begin{document}


\pagerange{\pageref{firstpage}--\pageref{lastpage}} \pubyear{2011}

\maketitle

\label{firstpage}

\begin{abstract}

We have mapped the stellar and gaseous kinematics, as well as the emission-line flux distributions and ratios, from the inner $\approx$\,450\,pc radius of the Seyfert 2 galaxy Mrk\,1157, using two-dimensional (2D) near-IR $J-$ and $K_l-$band spectra obtained with the Gemini NIFS instrument at a spatial resolution of $\approx$35~pc and velocity resolution of $\approx$\,40\,\kms.

The stellar velocity field shows a rotation pattern, with a discrete S-shaped zero velocity curve -- a signature of a nuclear bar. The presence of a bar is also supported by the residual map between the observed rotation field and a model of circular orbits in a Plummer potential. The stellar velocity dispersion ($\sigma_*$) map presents a partial ring of low-$\sigma_*$ values ($50-60$\,km\,s$^{-1}$) at 250\,pc from the nucleus surrounded by higher $\sigma_*$ values from the galaxy bulge. We propose that this ring has origin in kinematically colder regions with recent star formation. The velocity dispersion of the bulge (100\,\kms) implies in a black hole mass of $M_{BH}=8.3^{+3.2}_{-2.2}\times10^6$\,M$_\odot$.

Emission-line flux distributions are most extended along PA$=27/153^\circ$, reaching at least 450\,pc from the nucleus and following the orientation observed in previous optical emission-line \oiii\ imaging and radio jet. The molecular \h2\ gas has an excitation temperature $T_{\rm exc}\approx2300$\,K and its emission is dominated by thermal processes, mainly due to X-ray heating by the active nucleus, with a possible small contribution from shocks produced by the radio jet. The \feii\ excitation has a larger contribution from shocks produced by the radio jet, as evidenced by the line-ratio maps and velocity dispersion map, which show spatial correlation with the radio structures. The coronal lines are resolved, extending up to $\approx$\,150\,pc and are also slightly more extended along PA$=27/153^\circ$.

The gaseous kinematics shows two components, one due to gas located in the galaxy plane, in similar rotation to that of the stars and another in outflow, which is oriented close to the plane of the sky, thus extending to high latitudes, as the galaxy plane is inclined by $\approx$\,45$^\circ$ relative to the plane of the sky. The gas rotating in the plane dominates the \h2\ and \pb\ emission, while the gas in outflow is observed predominantly in  \feii\ emission. The \feii\  emission is originated in gas being pushed by the radio jet, which destroys dust grains releasing the Fe. From the outflow velocities and implied geometry, we estimate an outflow mass rate of  $\dot{M}_{\rm out}\approx6~{\rm M_\odot\, yr^{-1}}$ for the ionised gas and a kinetic power for the outflow of $\dot{E} \approx 2.3\times10^{41}$ erg\,s$^{-1}\,\approx 0.15\times L_{\rm bol}$.

The distinct flux distributions and kinematics of the \h2\ and \feii\ emitting gas, with the former more restricted to the plane of the galaxy, and the later tracing the outflows related to radio jets is a common characteristic of the 6 Seyfert galaxies (ESO\,428-G14, NGC\,4051, NGC\,7582, NGC\,4151, Mrk\,1066 and now Mrk\,1157) we have studied so far using similar 2D observations, and other 2 (Circinus and NGC\,2110) using long-slit observations. We conclude that the \h2\ emission surrounding the nucleus in the galaxy plane is a tracer of the gas feeding to the active nucleus while the \feii\ emission is a tracer of its feedback.  
\end{abstract}


\section{Introduction}

Recent resolved imaging and spectroscopic studies of the central regions of  nearby active galaxies have been allowing us to probe the feeding and feedback mechanisms of their central engines. The study of the of the ionized gas of the Narrow-Line Region (NLR) of active galaxies allows to investigate how the radiation and mass outflows from the nucleus interact with the circumnuclear gas, affecting its kinematics and excitation \citep[e.g.][]{fischer11,fischer10,crenshaw10a,crenshaw10b,crenshaw09,crenshaw07,kraemer09,holt06,veilleux05,schmitt96,veilleux97,wilson93}. On the other hand, the study of molecular and low ionization gas can tell us about the feeding of the Active Galactic Nucleus (AGN) of these galaxies by the mapping of inflows and estimating mass inflow rates towards the centre \citep[e.g.][]{vandeVen10,sanchez09,sb07,fathi06,mundell99}. However, most of the above studies are based on optical observations, which are affected by dust obscuration \citep{ferruit00,mulchaey96,mulchaey96b}, a problem that can be alleviated by using infrared lines to map the NLR emission.

Since 2006, we have been observing the inner few hundred parsecs of nearby active galaxies with Integral Field Spectrographs at the Gemini Observatory, with the goal of mapping both inflows and outflows around nearby active galactic nuclei and try to constrain the mass flow rates. In the near-infrared (hereafter near-IR) our main findings have been that the molecular (\h2) and ionised gases present distinct flux distributions and kinematics. The \h2\ emission gas is usually restricted to the plane of the galaxy, while the ionised gas extends also to high latitudes and is associated with the radio emission \citep{riffel06,n4051,n7582,mrk1066a,mrk1066c,sb09,sb10}. The \h2\ kinematics is usually dominated by rotation in the plane, including in some cases streaming motions towards the nucleus, while   
the kinematics of the ionised gas, and in particular of the \feii\ emitting gas, shows, in addition, a strong outflowing component associated with radio jets from the Active Galactic Nucleus (AGN). A previous study using long-slit near-IR data \citep{sb99} of the Seyfert galaxies Circinus and NGC\,2110, showing higher velocity dispersion and more disturbed kinematics for the \feii\ emission as compared with the \h2\ emission also supports this scenario. These results so far suggest that the molecular gas can be considered a tracer of the feeding of the AGN and the ionised gas a tracer of its feedback. Nevertheless, our 2D observations -- which provide a complete coverage of the kinematics -- comprise so far only half a dozen galaxies, and, in order to compensate for uncertain projection and filling factors, more objects need to be studied in detail, in order to allow also the quantification of the mass inflow and outflow rates.

In this work, we present the gaseous flux distribution and kinematics of the inner 450 pc of another active galaxy, the Seyfert~2 galaxy Mrk\,1157 (NGC\,591), for which we were able to measure also the stellar kinematics, allowing its comparison with the gas kinematics, which can be thus better constrained. This is only the second case in which we could measure the stellar kinematics, the first being Mrk\,1066 \citep{mrk1066c}. Mrk\,1157 was selected for this study because: (i) it presents strong near-IR emission lines \citep[e.g][]{nagar99,rogerio06}, allowing the mapping of the gaseous distribution and kinematics; (ii) it has extended radio emission, allowing the investigation of the role of the radio jet; (iii) the K-band CO absorption band heads have been observed in previous near-IR spectra \citep{rogerio06} allowing the measurement of the stellar kinematics. 

Mrk\,1157 is an early-type barred spiral galaxy (SB0/a), located at a distance $d=61.1$\,Mpc, for which 1\arcsec\ corresponds to 296\,pc at the galaxy. Radio-continuum images at 3.6~cm and 6~cm show a radio double with extension of 1\farcs2  oriented along the position angle PA=153$^\circ$ \citep{nagar99,ulvestad89}. Optical images show that the \oiii\ emission extends up to 4\arcsec\ following the orientation of the radio jet, while the \ha\ emission extends up to 20~\arcsec\ along the east-west direction \citep{mulchaey96}. The near-IR nuclear spectrum of Mrk\,1157 presents strong emission lines of [S\,{\sc iii}], He\,{\sc i}, H\,{\sc i}, \feii\ and \h2\ and the presence of high ionisation species as [Si\,{\sc vi}], [S\,{\sc viii}] and [Si\,{\sc x}] \citep[e.g.][]{rogerio06,veilleux97}. Optical and near-IR spectra show no evidence of broad components  in the profiles of permitted emission lines \citep[e.g.][]{rogerio06,veilleux97}, but spectropolarimetric observations reveal the presence of this component in \ha\ and \hb\ \citep{moran00}. The near-IR nuclear spectrum of Mrk\,1157 also shows absorption lines, with the CO absorptions in the H- and K-bands being the most prominent ones and including the detection of the CN absorption at 1.1$\mu$m -- a signature of the presence of intermediate age 
stars \citep{rogerio06,rogerio07}.

This paper is organised as follows. In Sec.~\ref{obs} we describe the observations and data reduction procedures. The results are presented in Sec.~\ref{results} and discussed in Sec.~\ref{discussion}. We present our conclusions in Sec.~\ref{conclusions}.

\section{Observations and Data Reduction}\label{obs}

\begin{figure*}
 \centering
 \includegraphics[scale=0.8]{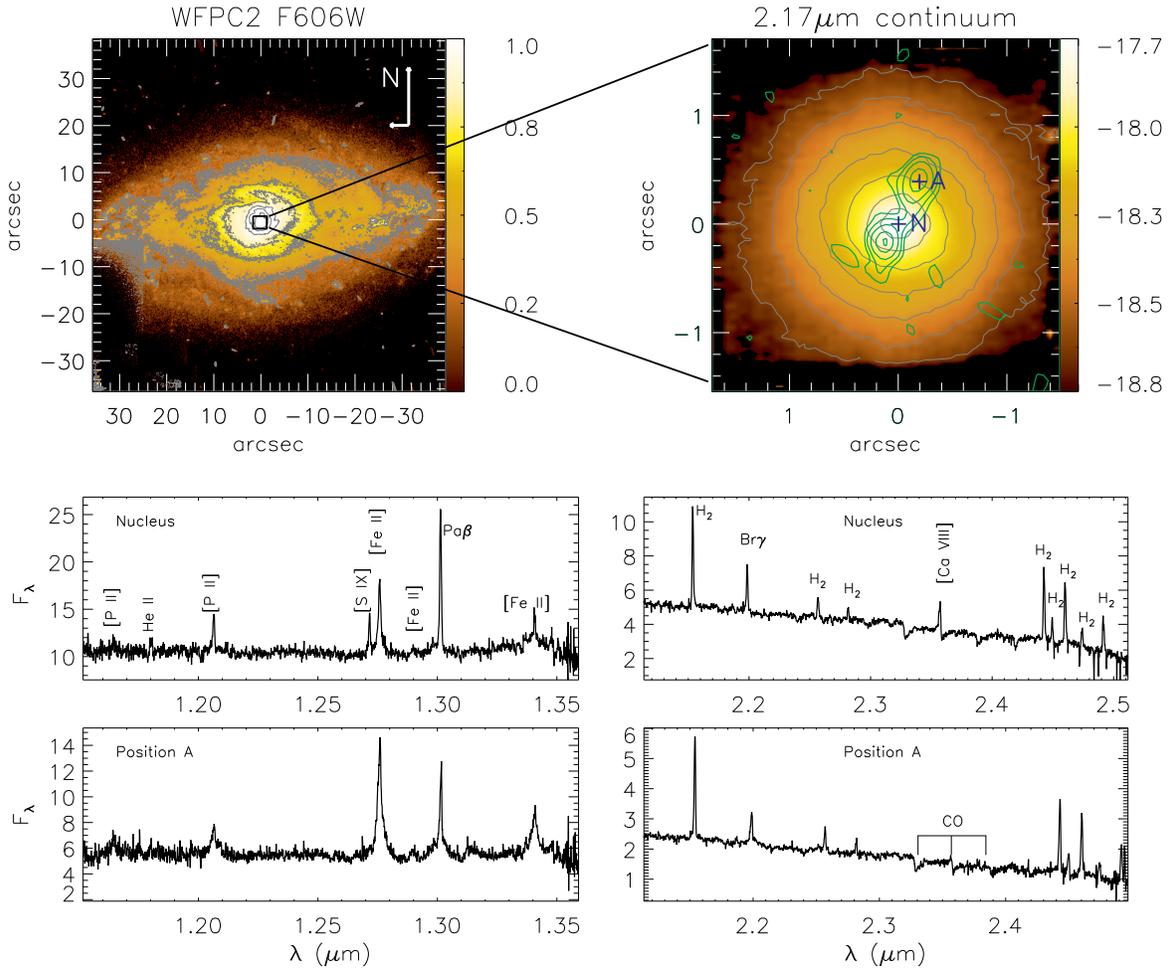} 
  \caption{Top-left panel: HST WFPC2 continuum image of Mrk\,1157 obtained through the filter F606W  \citep{malkan98}. Top-right panel: 2.17\,$\mu$m 
continuum image obtained from the NIFS data cube. The grey lines are flux contours from each image and the color bars show the flux scale in arbitrary units for the HST image and in logarithmic units for the NIFS image and the green contours are from the 3.6~cm radio-continuum image of \citet{nagar99}. Bottom panels show typical spectra obtained within an 0\farcs25$\times$0\farcs25 aperture for the nucleus and for a location at 0\farcs4 north-west from it (position A). The flux is shown in 10$^{-17}{\rm erg s^{-1} cm^{-2} \AA^{-1}}$ units.  The box in the HST image shows the NIFS field of view.} 
 \label{large}  
 \end{figure*}

Mrk\,1157 was observed with Gemini NIFS \citep{mcgregor03}  operating with the Gemini North Adaptive Optics system ALTAIR in September/October 2009 under the programme GN-2009B-Q-27, following the standard  Object-Sky-Sky-Object dither sequence, with off-source sky positions since the target is extended, and individual exposure times of 550\,s.

Two sets of observations with six on-source individual exposures were obtained at different spectral ranges: the first in the J-band, centred at 1.25\,$\mu$m and covering the spectral region from 1.14\,$\mu$m to 1.36\,$\mu$m, and the second in the K$_{\rm l}$-band, centred at 2.3\,$\mu$m  covering the spectral range from  2.10$\,\mu$m to 2.53$\,\mu$m.  In the J-band, the J\_G5603 grating and ZJ\_G0601 filter were used, resulting in a spectral resolution of $\approx1.8\,\AA$, as obtained from the measurement of the full width at half maximum (FWHM) of arc lamp lines. The K$_{\rm l}$-band observations were obtained using the Kl\_G5607 grating and HK\_G0603 filter and resulted in a spectral resolution of FWHM$\approx3.5\,\AA$. In velocity space, the resolution of the observations is $\approx$\,45\kms\ for the K$_{\rm l}$-band and 35 \kms\ for the J-band.

The data reduction was accomplished using tasks contained in the {\sc nifs} package which is part of {\sc Gemini iraf} package, as well as generic {\sc iraf} tasks. The reduction procedure included trimming of the images, flat-fielding, sky subtraction, wavelength and s-distortion calibrations. We have also removed the telluric bands and flux calibrated the frames by interpolating a black body function to the spectrum of the telluric standard star. The final IFU data cube in each band contains $\sim4500$ spectra, each spectrum corresponding to an angular coverage of 0$\farcs$05$\times$0$\farcs$05, which translates into $\sim$15$\times$15\,pc$^2$ at the galaxy and
covering the inner  3\arcsec$\times$3\arcsec ($\sim$900$\times$900\,pc$^2$) of the galaxy.

The angular resolution obtained  from the FWHM of the spatial profile of the telluric standard star is  0\farcs11$\pm$0\farcs02 for the J-band and 0\farcs12$\pm$0\farcs02 for the K$_{\rm l}$-band,  corresponding to 32.6$\pm$5.9 and 35.5$\pm$5.9\,pc at the galaxy, respectively.

\section{Results}\label{results}

In the top-left panel of Figure\,\ref{large} we present an optical image of Mrk\,1157 obtained with the Hubble Space Telescope (HST)  Wide Field Planetary Camera 2 (WFPC2) through the filter F606W 
 -- effective wavelength/width 5843\,\AA/1578.7\,\AA
 \citep{malkan98}. In the top-right panel we present an image obtained from the NIFS data cube for the continuum emission around 2.17$\,\mu$m.  In the bottom panels we present two characteristic IFU spectra integrated within a 0\farcs25$\times$0\farcs25 aperture: the nuclear one and a spectrum from 0\farcs4 north-west of the nucleus (Position A), at the location of the north-west component of the radio double \citep{nagar99,ulvestad89}.  


\subsection{Stellar Kinematics}\label{stellar_kin}

\begin{figure*}
 \centering
 \includegraphics[scale=0.8]{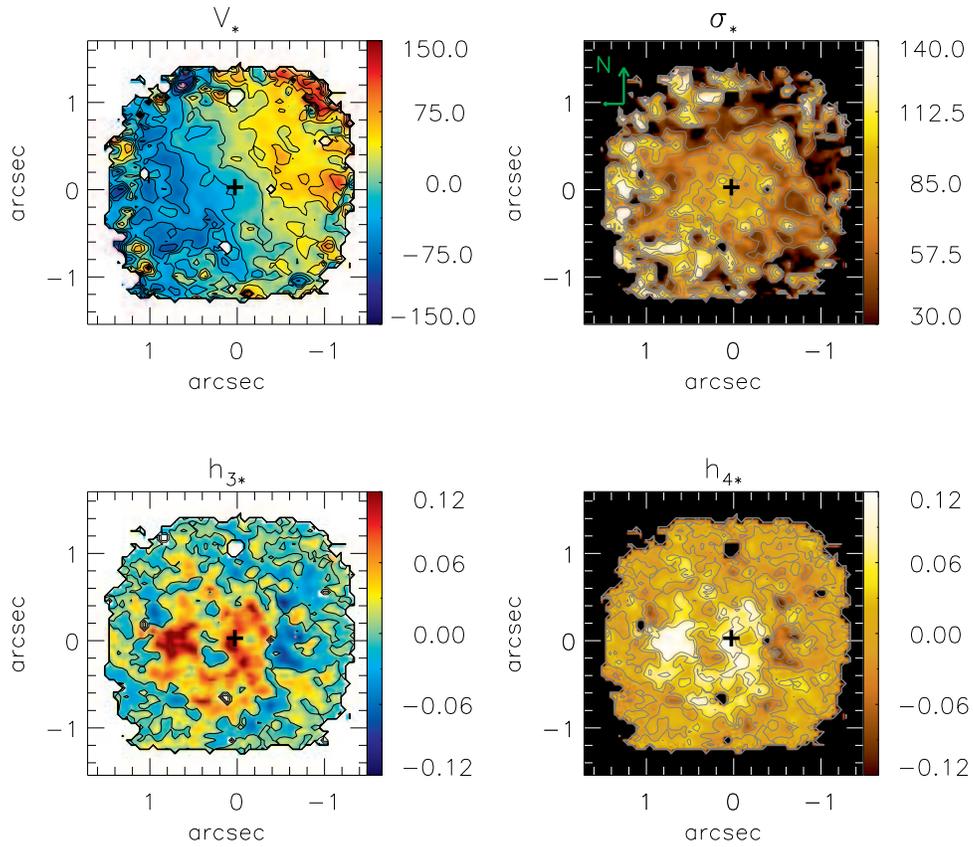} 
  \caption{Stellar kinematics obtained from the fitting of the CO band heads. Top left: centroid velocity field; top right: velocity dispersion map; bottom left: h$_{3*}$ map and bottom right: h$_{4*}$ map.  The colour bars show the range of values for the velocity and velocity dispersion in km\,s$^{-1}$ (top) and for h$_3$ and h$_4$ Gauss-Hermite moments (bottom).} 
 \label{stel}  
 \end{figure*}

In order to obtain the stellar line-of-sight velocity distributions (LOSVD) we fitted the $^{12}$CO and $^{13}$CO  stellar absorption band heads around 2.3\,$\mu$m in the $K$-band spectra using the penalised Pixel-Fitting ({\sc ppxf}) method of \citet{cappellari04}, as explained in \citet{mrk1066c}. As stellar template spectra we used those of the Gemini library of late spectral type stars observed with the Gemini Near-Infrared Spectrograph (GNIRS) IFU and  NIFS \citep{winge09}. The resulting maps for the radial velocity (V$_*$), stellar velocity dispersion ($\sigma_*$), and higher order Gauss-Hermite moments ($h_{3*}$ and  $h_{4*}$) are shown in Figure~\ref{stel}. At locations close to the borders of the field of view, the S/N ratio of the galaxy's spectra was not high enough to allow reliable measurements, which have thus been masked out.

In order to estimate the uncertainties on the measurements of $V_*$, $\sigma_*$, $h_{3*}$ and $h_{4*}$ we performed 500 iterations Monte Carlo simulations of the kinematical extractions with  pPXF,using the Gemini library spectra as templates, as described in \citet{cappellari04}. Our NIFS spectra have S/N ratio  in the 2.3$\,\mu$m continuum ranging from 15 at the borders of the NIFS field to 90 around the nucleus.  We have performed two Monte Carlo simulations, one for ${\rm S/N=20}$ to obtain the uncertainties near the borders of the field and another for ${\rm S/N=60}$ to represent the uncertainties at regions near the nucleus. The output from the Monte Carlo simulations 
 show that for a ${\rm S/N=60}$ the maximum uncertainties are  $\Delta V_*\approx5$\,\kms, $\Delta \sigma_*\approx5$\,\kms, $\Delta h_{3*}\approx0.04$ and $\Delta h_{4*}\approx0.04$, while for ${\rm S/N=20}$ they are $\Delta V_*\approx15$\,\kms, $\Delta \sigma_*\approx20$\,\kms, $\Delta h_{3*}\approx0.05$ and $\Delta h_{4*}\approx0.04$.




The top left panel of Fig.\,\ref{stel} shows the stellar velocity field, from which we subtracted the heliocentric systemic velocity of $V_s=4473\pm8\,{\rm km\,s^{-1}}$, obtained from the modelling of the  stellar velocity field (see Sec.~\ref{disc-stel}).  The velocity field shows a rotation pattern with blueshifts to the south-east and redshifts to the north-west, with a maximum velocity of $\approx$130\,\kms\ and kinematical centre coincident with the position of the continuum peak flux, within the uncertainties.
 
In the top-right panel of Fig.~\ref{stel} we present the stellar velocity dispersion ($\sigma_*$) map, which presents values ranging from 30 to 150~\kms. A partial ring  of low $\sigma_*$ values ($\approx50\,{\rm km\,s^{-1}}$) is observed surrounding the nucleus at 0\farcs6 ($\approx$\,180\,pc) from it. The bottom panels  show the higher order Gauss-Hermite moments $h_{3*}$ (left) and $h_{4*}$ (right), which measure asymmetric and symmetric deviations, respectively, from a Gaussian velocity distribution. The values $h_{3*}$ and $h_{4*}$ are small, varying from $-$0.12 to 0.12, and are similar to those observed for other galaxies \citep{emsellem04,ganda06,n4051,n7582}, indicating that there are only small deviations of the stellar LOSVD from a Gaussian velocity distribution. Nevertheless, some systematic deviations are observed in both maps. The  $h_{3*}$ seems to be anti-correlated with the stellar velocity field, presenting red wings  (positive values) mostly in the blueshifted side of the galaxy, and blue wings  (negative values) in the redshifted side. The $h_{4*}$ map has the highest positive values  (meaning the the LOSVD is more ``pointy"  than a Gaussian) where the red wings are observed and the highest negative values (meaning that the LOSVD is less ``pointy"  than a Gaussian) where the blue wings are observed.


\subsection{Emission-Line Flux Distributions and Ratios}\label{flux_distributions}

We measured the fluxes of 29 emission lines from [P {\sc ii}], [Fe\,{\sc ii}], He\,{\sc i}, He\,{\sc ii}, H\,{\sc i}, \h2, [S\,{\sc ix}], [Si\,{\sc vii}] and [Ca\,{\sc viii}], which are listed in  Table\,\ref{fluxes} for the nucleus and position A. The strongest emission lines are identified in the nuclear spectrum of Fig.\,\ref{large}. In the K-band nuclear spectrum we have also identified the CO stellar absorption band heads around 2.3\,$\mu$m.

\begin{table*}
\centering
\caption{Measured emission-line fluxes (in units of 10$^{-17}$\,erg\,s$^{-1}$\,cm$^{-2}$) for the two positions marked in Fig.~\ref{large} within 0\farcs25$\times$0\farcs25 aperture.}
\vspace{0.3cm}
\begin{tabular}{l l c c}
\hline
$\lambda_{vac} {\rm(\mu m)}$   & ID                     & Nucleus               & Position A         \\
\hline
1.14713  & [P {\sc ii}]\,$^1D_3-^3P_1$                &  16.39 $\pm$ 1.07     &  21.37 $\pm$ 7.20  \\
1.16296  & He\,{\sc ii}\,$7-5$                        &  19.84 $\pm$ 4.01     &   7.71 $\pm$ 0.78  \\
1.18861  & [P {\sc ii}]\,$^1D_2-^3P_2$                &  38.48 $\pm$ 3.49     &  49.68 $\pm$ 8.03  \\
1.19665  & [Fe\,{\sc ii}]\,$b^4D_{5/2}-a^6F_{3/2}$    &   6.00 $\pm$ 0.82     &   2.48 $\pm$ 0.33  \\
 1.19723 &  He\,{\sc i}\,$^3D_1-^3Po_{2}$             & 	     --       &   1.79 $\pm$ 0.50  \\
1.22263  & [Fe\,{\sc ii}]\,$a^4D_{1/2}-a^6D_{5/2}$    &   2.53 $\pm$ 0.27     &  --		\\
1.23878  & [Fe\,{\sc ii}]\,$c^2G_{7/2}-a^4G_{11/2}$   &   4.64 $\pm$ 0.67     &    5.19 $\pm$ 0.91  \\
1.24211  & [Fe\,{\sc ii}]\,$b^4D_{1/2}-a^6S_{5/2}$    &  --	              &    2.13 $\pm$ 1.15  \\
1.25235  & [S\,{\sc ix}]\,$^3P_1-^3P_2$               &  34.83 $\pm$ 1.57     &    4.57 $\pm$ 1.17  \\
1.25702  & [Fe\,{\sc ii}]\,$a^4D_{7/2}-a^6D_{9/2}$    & 101.07 $\pm$ 7.65     &  207.82 $\pm$ 19.66  \\
1.27069  & [Fe\,{\sc ii}]\,$a^4D_{1/2}-a^6D_{1/2}$    &  --	              &   18.07 $\pm$ 3.23  \\
1.27912  & [Fe\,{\sc ii}]\,$a^4D_{3/2}-a^6D_{3/2}$    &   7.18 $\pm$ 1.31     & --		    \\       
1.28216  &  H\,{\sc i}\,Pa$\beta$                     & 123.27 $\pm$ 3.05     &  101.69 $\pm$ 31.30  \\
1.28607  & [Fe\,{\sc ii}]\,$c^2D_{5/2}-b^6G_{9/2}$    & 	       --     &    1.84 $\pm$ 0.29  \\
1.29462  & [Fe\,{\sc ii}]\,$a^4D_{5/2}-a^6D_{5/2}$    &   --	              &    5.85 $\pm$ 0.69  \\
1.32092  & [Fe\,{\sc ii}]\,$a^4D_{7/2}-a^6D_{7/2}$    &  32.76 $\pm$ 3.15     &   67.50 $\pm$ 8.36  \\
1.32814  & [Fe\,{\sc ii}]\,$a^4D_{5/2}-a^6D_{3/2}$    &  13.99 $\pm$ 5.80     &   8.77 $\pm$ 7.4  \\

2.12183  & H$_2$\, 1-0\,S(1) 	  		      &     74.57 $\pm$  9.48 & 50.46 $\pm$  2.43  \\
2.15420  &  H$_2$\,1-0\,S(2)  	                      &      3.48 $\pm$  3.60 &  2.70 $\pm$  0.96  \\
2.16612  & H\,{\sc i}\,Br$\gamma$                     &     46.14 $\pm$  6.33 & 19.63 $\pm$  1.53  \\
2.22344  &  H$_2$\,1-0\,S(0)  	                      &     22.32 $\pm$  0.90 &  13.02 $\pm$  1.08 \\
2.24776  &  H$_2$\,2-1\,S(1)  	                      &     8.41 $\pm$  3.43 &   6.27 $\pm$  1.36 \\
2.32204  &[Ca\,{\sc viii}]\,$^2P^0_{3/2}-^2P^0_{1/2}$ &     36.83 $\pm$ 11.25 &   5.12 $\pm$  0.99 \\
2.40847  &  H$_2$\,1-0\,Q(1)   	                      &     56.59 $\pm$  2.69 &  33.08 $\pm$  0.78 \\
2.41367  &  H$_2$\,1-0\,Q(2)   	                      &     14.46 $\pm$  6.18 &  10.23 $\pm$  1.28 \\
2.42180  &  H$_2$\,1-0\,Q(3)   	                      &     51.63 $\pm$  3.48 &  33.77 $\pm$  2.02 \\
2.43697  &  H$_2$\,1-0\,Q(4)   	                      &     16.21 $\pm$  3.88 &   9.03 $\pm$  1.59 \\
2.45485  & H$_2$\,1-0\,Q(5)   	                      &     32.20 $\pm$  3.52 &  19.82 $\pm$  4.39 \\
2.48334  & [Si\,{\sc vii}] $^3P_1 - ^3P_2$            &     85.95 $\pm$  5.64 &  17.88 $\pm$  1.36 \\

\hline

\end{tabular}
\label{fluxes}
\end{table*}

\begin{figure*}
 \centering
 \includegraphics[scale=0.8]{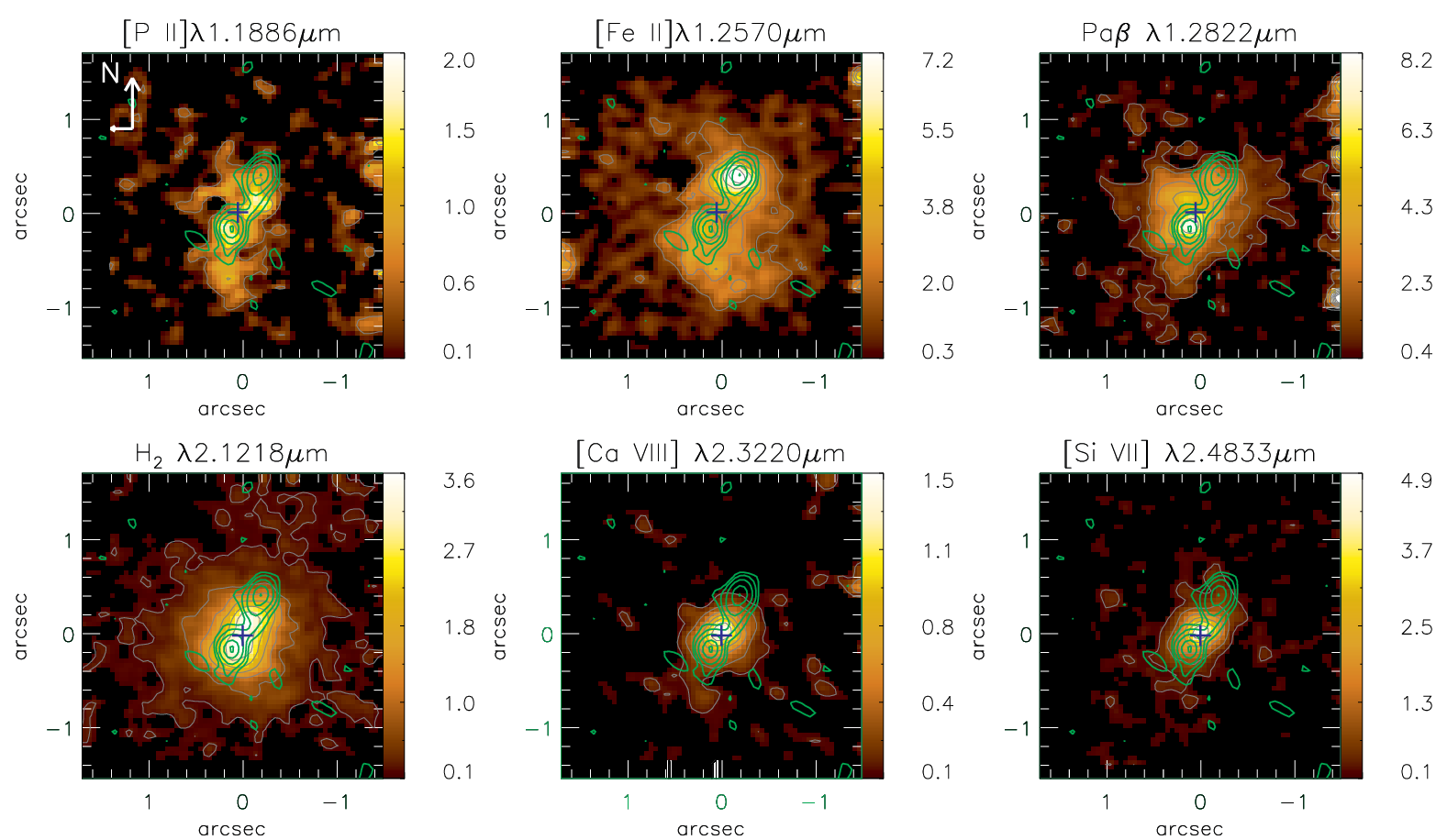} 
  \caption{Emission-line flux distributions 
The central cross marks the position of the nucleus, the grey contours correspond to each flux map, while the green contours are from the 3.6~cm radio-continuum image of \citet{nagar99}. Colour bars show the range of flux values for each emission line in units of 10$^{-17}\,{\rm erg\,s^{-1}\,cm^{-2}}$.}  
 \label{flux}  
 \end{figure*}
 
We used our routine {\sc profit} \citep{profit} to fit the profiles of [P\,{\sc ii}]\,$\lambda$1.1886\,$\mu$m, [Fe\,{\sc ii}]\,$\lambda$1.2570\,$\mu$m, Pa$\beta$, H$_2\,\lambda$2.1218$\mu$m, [Ca\,{\sc viii}]\,$\lambda$2.3220\,$\mu$m and [Si\,{\sc vii}]\,$\lambda$2.4833\,$\mu$m emission lines with Gauss-Hermite series. We have integrated the flux under the emission-line profiles after subtracting the underlying continuum (obtained  from two spectral windows, one at each side of the profile) in order to map of the flux distributions. These particular lines have been chosen because they have the highest signal-to-noise (S/N) ratios among their species (coronal lines, forbidden and permitted  ionised gas lines and molecular lines). As observed in the nuclear spectrum of  Fig.~\ref{large}, the [Ca\,{\sc viii}] emission line is partially superimposed on  the $^{12}$CO\,$3-1\,\lambda$2.322\,$\mu$m band head. In order to correct for the effect of the absorption, we measured the flux of [Ca\,{\sc viii}] emission line after the subtraction of the contribution of the underlying stellar population, obtained  from the fitting of the stellar kinematics (see Sec.~\ref{stellar_kin}). 

In Figure~\ref{flux} we present the resulting flux distributions, which have typical uncertainties  smaller than 10\%. The central cross marks the position of the nucleus, defined as the location of the peak of the continuum emission,  and the green contours overlaid on the  [Fe\,{\sc ii}] and Pa$\beta$ flux maps are from the 3.6~cm radio-continuum image from \citet{nagar99}. We used the 20~cm radio-continuum image from \citet{nagar99} to align the near-IR and radio images, under the assumption that both present their peak emission at the same position -- the nucleus. The uncertainty in the alignment is estimated to be smaller than 0\farcs2. 

The flux distributions for the [P\,{\sc ii}] and [Fe\,{\sc ii}] emission lines are similar. They are well correlated with the 3.6~cm radio-continuum emission, being more extended along the position angle of the radio double PA=153$\degr$. The peak flux is observed at 0\farcs4 north-west of the nucleus, approximately coincident with the north-west component of the radio double. A secondary peak is observed at 0\farcs2 south-east of the nucleus, at the same position of the other component of the radio double (hereafter we will call the radio components NW and SE hot spots). 

The \pb\ emission is also extended along PA=153$^\circ$, but shows some emission also to the east of the nucleus, being less collimated than the [P\,{\sc ii}] and [Fe\,{\sc ii}]  flux distributions; its emission peaks at 0\farcs2 south-east of the nucleus, at the position of the SE hot spot. 

The \h2\ emission is more uniformly distributed, extending up to 0\farcs8 arcsec along all directions, but the highest intensity levels are also more elongated along the orientation of the radio double. 

The coronal-line emission (traced by the [Ca\,{\sc viii}] and [Si\,{\sc vii}] emission lines) is resolved by our observations (the FWHM of the flux distribution for both emission lines is $\approx$0\farcs4), peaks at the nucleus and extends up to 0\farcs4 ($\approx$120\,pc) from it. The [Si\,{\sc vii}] coronal-line emission in more extended along PA=153$^\circ$, as observed for the other emission lines.

\subsubsection{Emission-line Ratios}

\begin{figure*}
 \centering
 \includegraphics[scale=0.8]{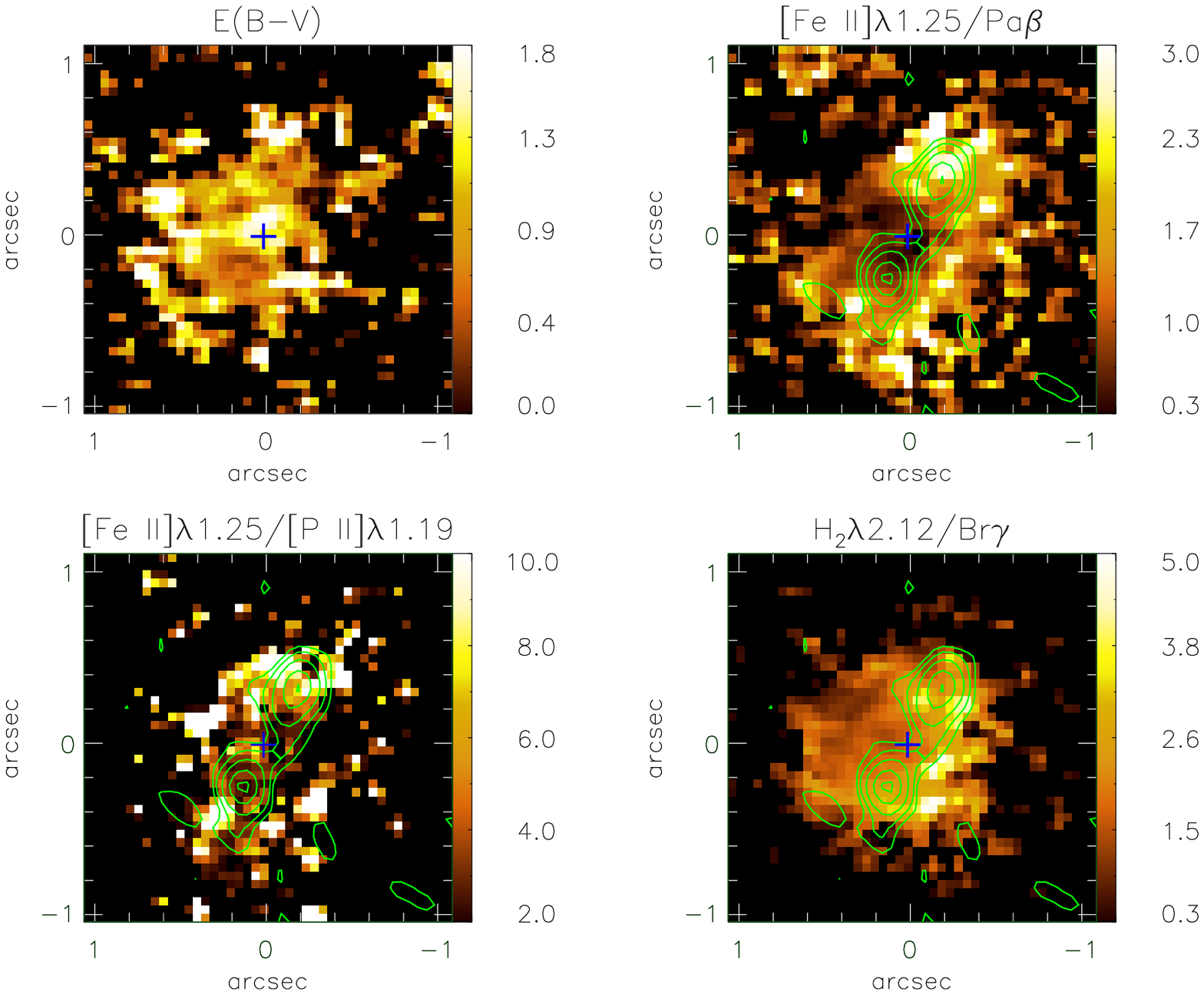} 
  \caption{Emission-line ratio maps; the reddening E(B-V)  was obtained from Pa$\beta$/Br$\gamma$ line ratio (top-left panel). The central cross marks the position of the nucleus and the green contours are from the 3.6~cm radio-continuum image of \citet{nagar99}. We do not show the borders of field (the outermost 0\farcs5) due to the small S/N ratio in the line-ratio maps.
 } 
 \label{ratio}  
 \end{figure*}

In order to map the Narrow-Line Region (NLR) extinction and excitation mechanisms of the [Fe\,{\sc ii}] and H$_2$ emission lines, we constructed the flux ratio maps shown in Figure\,\ref{ratio}. 

The reddening map was obtained from the Pa$\beta$/Br$\gamma$ line ratio as 
\begin{equation}
 E(B-V)=4.74\,{\rm log}\left(\frac{5.88}{F_{Pa\beta}/F_{Br\gamma}}\right),
\end{equation}
where $F_{Pa\beta}$ and $F_{Br\gamma}$ are the fluxes of $Pa\beta$ and $Br\gamma$ emission lines, respectively. We have used the reddening law of \citet{cardelli89} and adopted the intrinsic ratio $F_{Pa\beta}/F_{Br\gamma}=5.88$ corresponding to case B recombination \citep{osterbrock06}. The resulting $E(B-V)$ map is shown in the top-left panel of Fig.\,\ref{ratio},  where the highest $E(B-V)$ values -- corresponding to the most heavily reddened regions -- are those represented by the colours light yellow and white. The highest values, of up to 1.8, are observed at the nucleus and to north-east, while the lowest values, of about 0.4, are observed at the radio hot spots. Typical uncertainties are smaller than 0.2, and the average value is $E(B-V)\approx0.5$.

In the top-right panel of Fig.\,\ref{ratio} we present the [Fe\,{\sc ii}]$\lambda$1.2570$\,\mu$m/Pa$\beta$ ratio map, which can be used to investigate the excitation mechanism of  [Fe\,{\sc ii}] \citep[e.g.][]{ardila04,ardila05,sb09,riffel06,mrk1066a,sb09}.
   Seyfert galaxies present typical values for this ratio between 0.6 and 2.0, Starbursts have ratios $\lesssim$0.6 and Low-Ionisation Nuclear Emission-line Regions (LINERs) show values higher than 2.0 \citep[e.g.][]{ardila04,ardila05,sb09,riffel06,mrk1066a,sb09}. 
The lowest values for Mrk\,1157 ($\approx0.3$)  are observed near the nucleus (closer than $\approx$\,0\farcs4 or 120\,pc from the nucleus), while the highest values, of up to 3, are observed at $\approx$\,0\farcs4 north-west of the nucleus, at the position of the NW radio hot spot. 

Another line ratio that can be used to investigate the [Fe\,{\sc ii}] excitation mechanism is 
[Fe\,{\sc ii}]$\lambda$1.2570$\,\mu$m/[P\,{\sc ii}]$\lambda$1.8861$\,\mu$m.  Values larger than
2 indicate that shocks have passed through the gas destroying the dust grains, releasing the Fe and enhancing its abundance and thus emission  \citep[e.g.][]{oliva01,sb09,mrk1066a}. We present this ratio map in the bottom-left panel of Fig.\,\ref{ratio}. The lowest values of $\approx2$ are observed at the nucleus, while the highest values of up to 10 are seen at 0\farcs4 north-west, at the location of the NW hot spot.

In the bottom-right panel of Fig.\,\ref{ratio} we present  the H$_2\lambda$2.1218$\,\mu$m/Br$\gamma$ ratio map, which is useful to investigate the excitation of the H$_2$ molecule \citep[e.g.][]{ardila04,ardila05,sb09,riffel06,n4051,n7582,mrk1066a}. 
This ratio presents values ranging from 0.6 to 2 for Seyfert galaxies
\citep[e.g.][]{ardila04,ardila05,sb09,riffel06,n4051,n7582,mrk1066a}.
The lowest values, down to $\approx$\,0.3 are observed to the north-east of the nucleus and the highest 
values of up to 5 are observed mostly along a strip running from  south-west to north-west of the nucleus.

\subsection{Gas kinematics}

The {\sc profit} routine \citep{profit}, which we have used to fit the emission line profiles, also outputs  the centroid velocity ($V$), velocity dispersion ($\sigma$) and higher order Gauss-Hermite moments ($h_3$ and $h_4$), which have been used to map the gas kinematics.

\begin{figure*}
 \centering
 \includegraphics[scale=0.84]{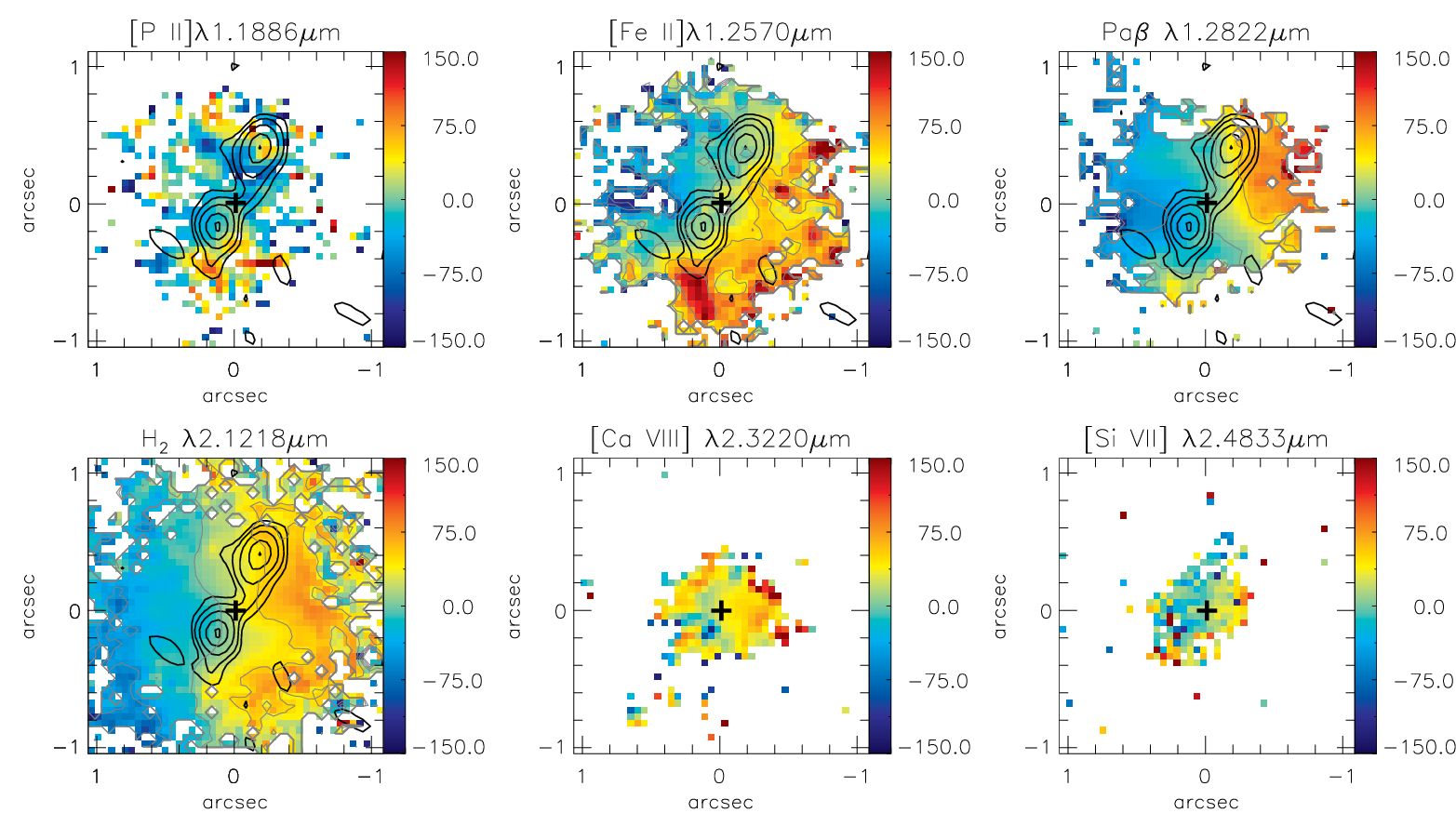} 
  \caption{Centroid velocity fields for the 
 [P\,{\sc ii}]\,$\lambda$1.1886\,$\mu$m (top left), [Fe\,{\sc ii}]\,$\lambda$1.2570\,$\mu$m (top central), Pa$\beta$ (top right), H$_2\,\lambda$2.1218$\mu$m (bottom left), [Ca\,{\sc viii}]\,$\lambda$2.3220\,$\mu$m (bottom central) and [Si\,{\sc vii}]\,$\lambda$2.4833\,$\mu$m (bottom right) emitting gas.  The central cross marks the position of the nucleus and the black contours are from the 3.6~cm radio-continuum image of \citet{nagar99}. We do not show the outermost  0\farcs5 borders of the NIFS field, since it was not possible to fit the emission line profiles at these locations due to the small S/N ratio. The color bars show the range of values for the velocities in km\,s$^{-1}$. } 
 \label{vel}  
 \end{figure*}

In Figure~\ref{vel} we present the velocity fields obtained from the centroid wavelength of each emission line, with typical mean uncertainty of 8 \kms. White regions in this figure represent locations where the S/N was not high enough to allow the fitting of the line profiles. Most velocity fields show mostly blueshifts to the east (left in the figures) and redshifts to the west, with the line of zero velocity running approximately vertically (north-south) for \pb\ and \h2\ and from the north-west to the south-east for \feii. The gas velocity fields differ from the stellar one, for which the line of zero velocity runs from the north-east to the south-west. The \feii\ emitting gas presents higher redshifts than the other emission lines, of up to 150\,\kms, observed to south-east  of  the nucleus, near to the edge of the radio structure. Similar redshifts are also observed in the \pii\ velocity field, which presents additionally some high blueshifts. As observed in the bottom-central and bottom-right panels, the coronal emitting gas, although just barely extended, seems to show similar kinematics to that observed for the rest of the ionized emitting gas.

\begin{figure*}
 \centering
 \includegraphics[scale=0.84]{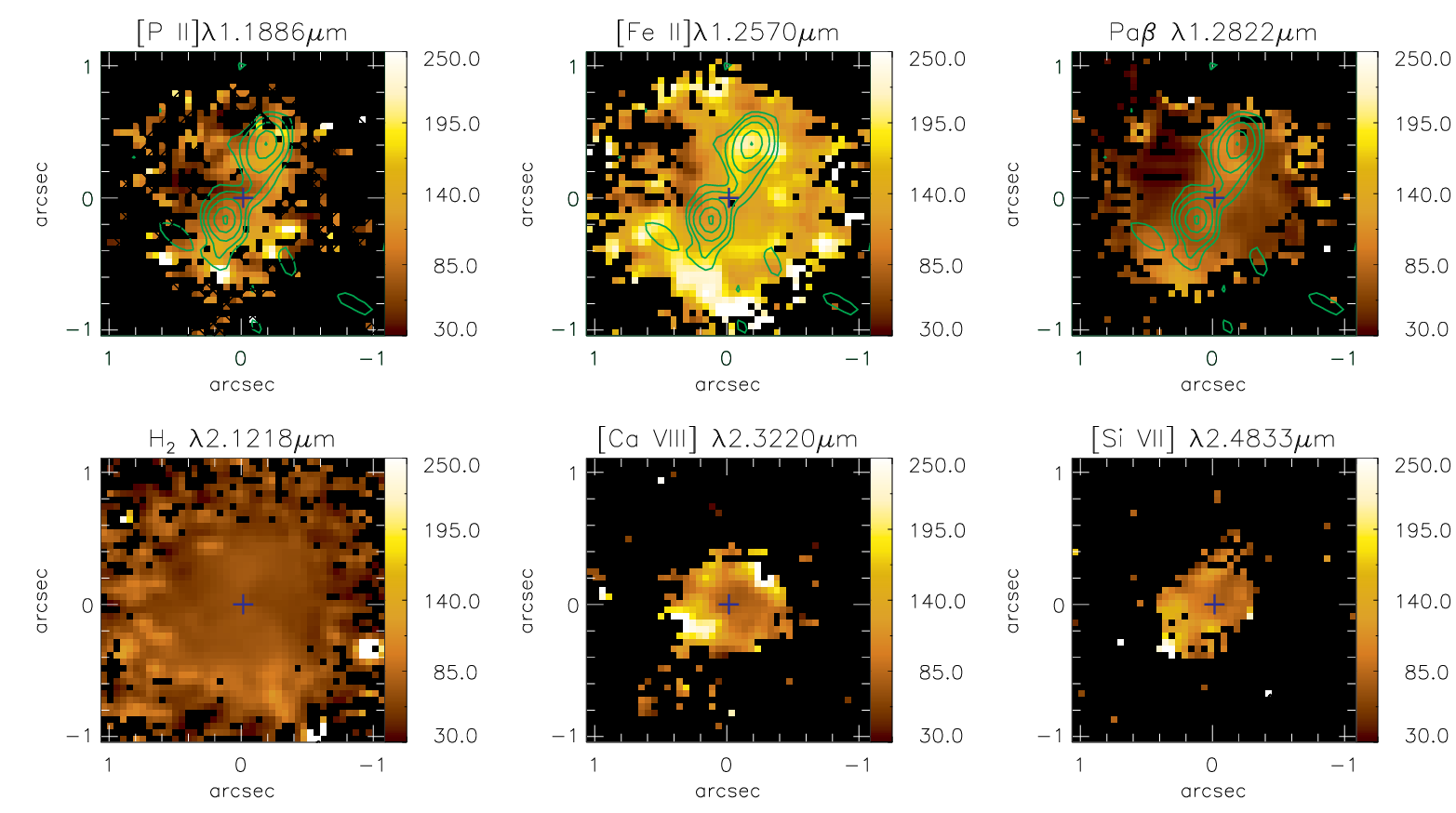} 
  \caption{$\sigma$ maps for the same emission lines of Fig.\ref{vel}.  The central cross marks the position of the nucleus and the green contours are from the 3.6~cm radio-continuum image of \citet{nagar99}. We do not show the outermost 0\farcs5 borders of the NIFS field, since it was not possible to fit the emission line profiles at these locations due to small S/N ratio. The color bars show the range of $\sigma$ values in km\,s$^{-1}$.}
 \label{sig}  
 \end{figure*}

\begin{figure*}
 \centering
 \includegraphics[scale=0.84]{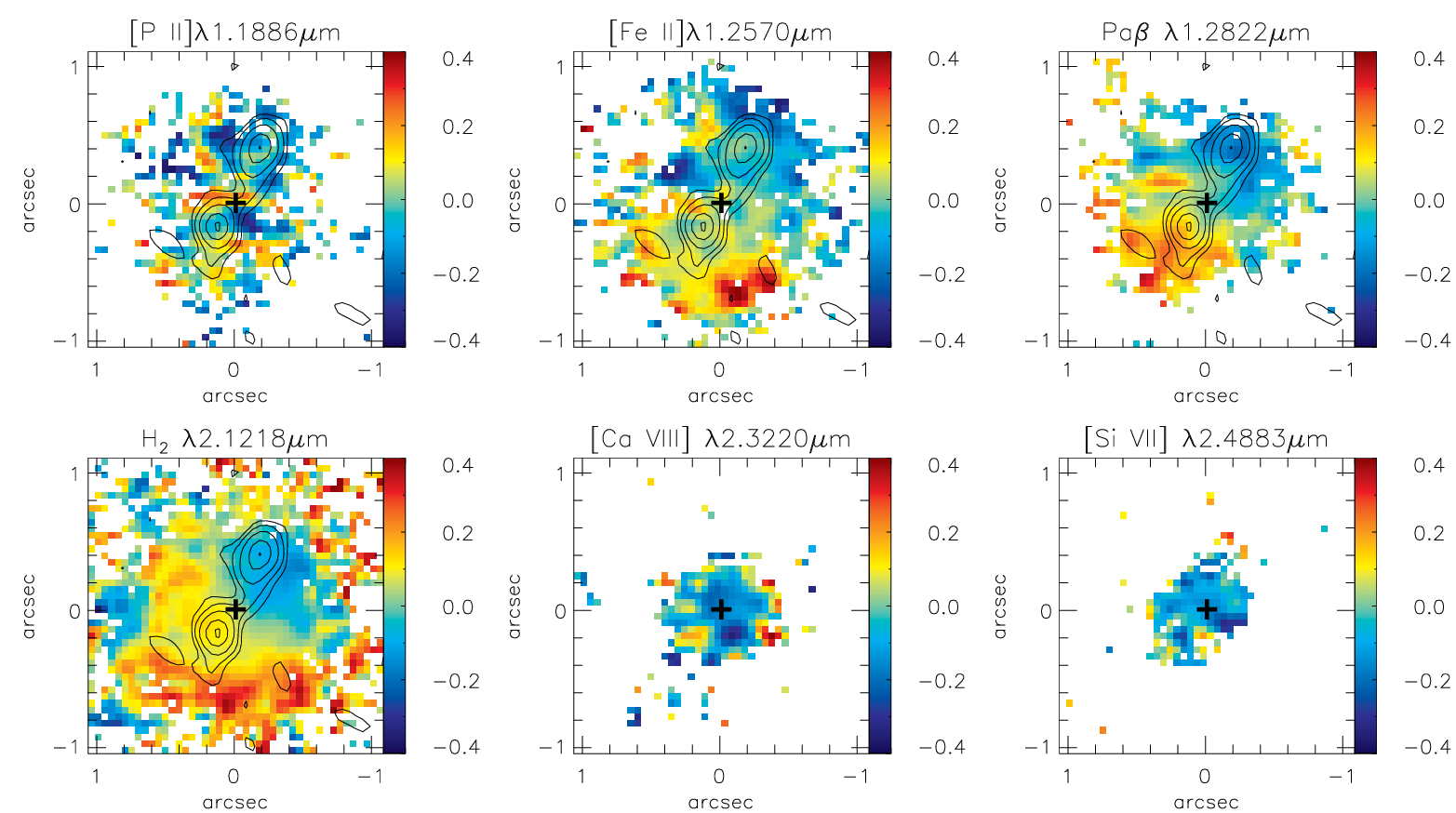} 
  \caption{$h_3$ maps the same emission lines of Fig.\ref{vel}. The central cross marks the position of the nucleus and the contours are from the 3.6~cm radio-continuum image of \citet{nagar99}. We do not show the outermost  0\farcs5 borders of the NIFS field, since it was not possible to fit the emission line profiles at these locations due to small S/N ratio. The color bars show the range of $h_3$ values.} 
 \label{h3}  
 \end{figure*}


Figure~\ref{sig} shows the velocity dispersion maps, which have typical mean uncertainties of 10 \kms. The green contours overlaid to the \feii\ $\sigma$ map are from the 3.6~cm radio continuum image. The \feii~$\sigma$ map shows the highest values of up to $\approx$250~\kms~in regions at and around the radio hot spots, while the lowest values ($\approx70$~\kms) are observed predominantly to north-east of the nucleus, away from the radio structure. The \pii~$\sigma$ map is similar to the \feii~$\sigma$ map, although with $\approx$30\% lower values. The same is observed in the  \pb~$\sigma$ map but with even lower values (50\% lower), ranging from  $\approx$50~\kms  (observed to north-east of the nucleus) to $\approx$\,130~\kms, observed along PA=153$\degr$ in association with the radio structure. The \h2\ emitting gas presents the lowest  $\sigma$ values, which are smaller than 60~\kms\ at most locations.  The coronal lines have similar  $\sigma$ values to those observed for the \pii, with median values of $\approx$110~\kms, being smaller than those seen in the \feii. 

In Figure~\ref{h3} we present the $h_3$ maps, which show values ranging from $-0.4$ to $0.4$ for most emission lines. The mean uncertainties in $h_3$ are smaller than 0.03 for all emission lines. The $h_3$ maps for the \pii, \feii, \pb\ and \h2\ emission lines show negative values -- meaning blue wings -- in the profiles to the north-west and positive values -- meaning red wings -- to the south-east, with the highest absolute values observed near the borders of the radio structure. The coronal lines of [Ca\,{\sc viii}] and [Si\,{\sc vii}] present mostly negative values, indicating the dominance of blue wings in their profiles. 

The $h_4$ maps show values close to zero at most locations and are thus not shown. The mean uncertainties are smaller than 0.05 for all emission lines. The \feii\ and \pii\ show a few high values, of up to 0.4, to north-east and to south-west -- away from the radio structure, meaning that the profiles are more peaked than gaussians at these locations.

\subsection{Channel Maps}

\begin{figure*}
 \centering
 \includegraphics[scale=0.67]{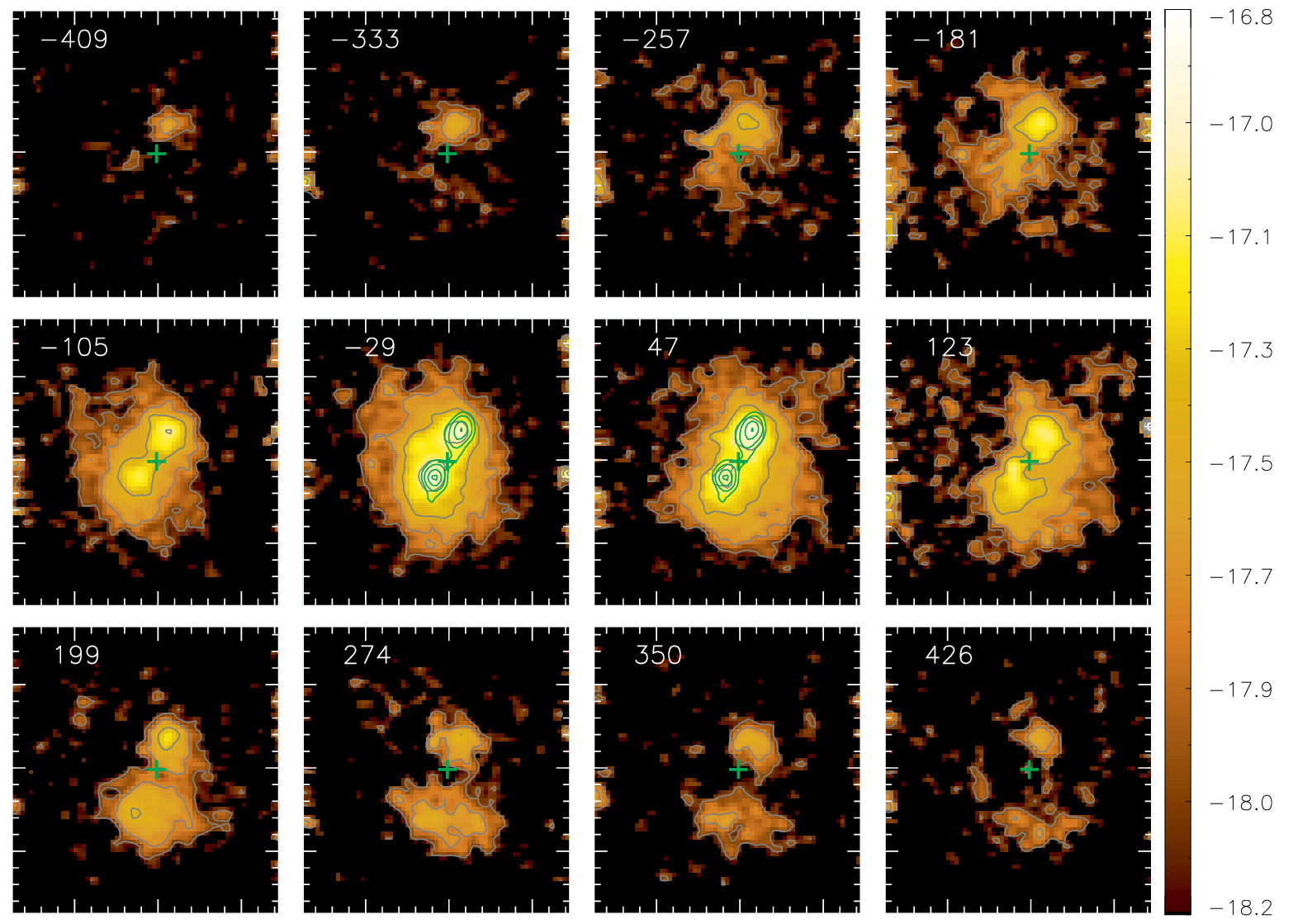} 
  \caption{Channel maps along the \feii\ emission-line profile for a velocity bin of 76\,\kms\ and centred at the velocities shown in the top-left corner of each panel.  The green contours are from the 3.6~cm radio image of \citet{nagar99} and the cross marks the position of the nucleus. The color bar shows the flux scale in logarithmic units.}  
 \label{slice_fe}  
 \end{figure*}

\begin{figure*}
 \centering
 \includegraphics[scale=0.67]{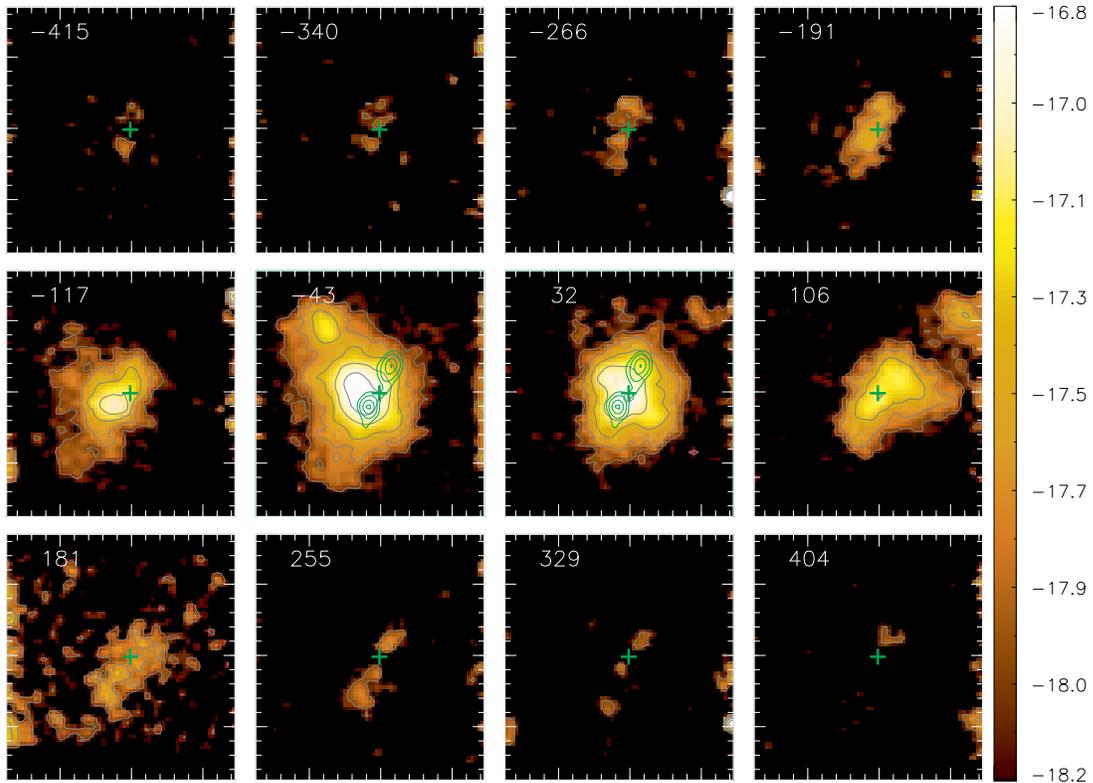} 
  \caption{Same as Fig.~\ref{slice_fe} for the  \pb\ emission line.} 
 \label{slice_pb}  
 \end{figure*}

\begin{figure*}
 \centering
 \includegraphics[scale=0.67]{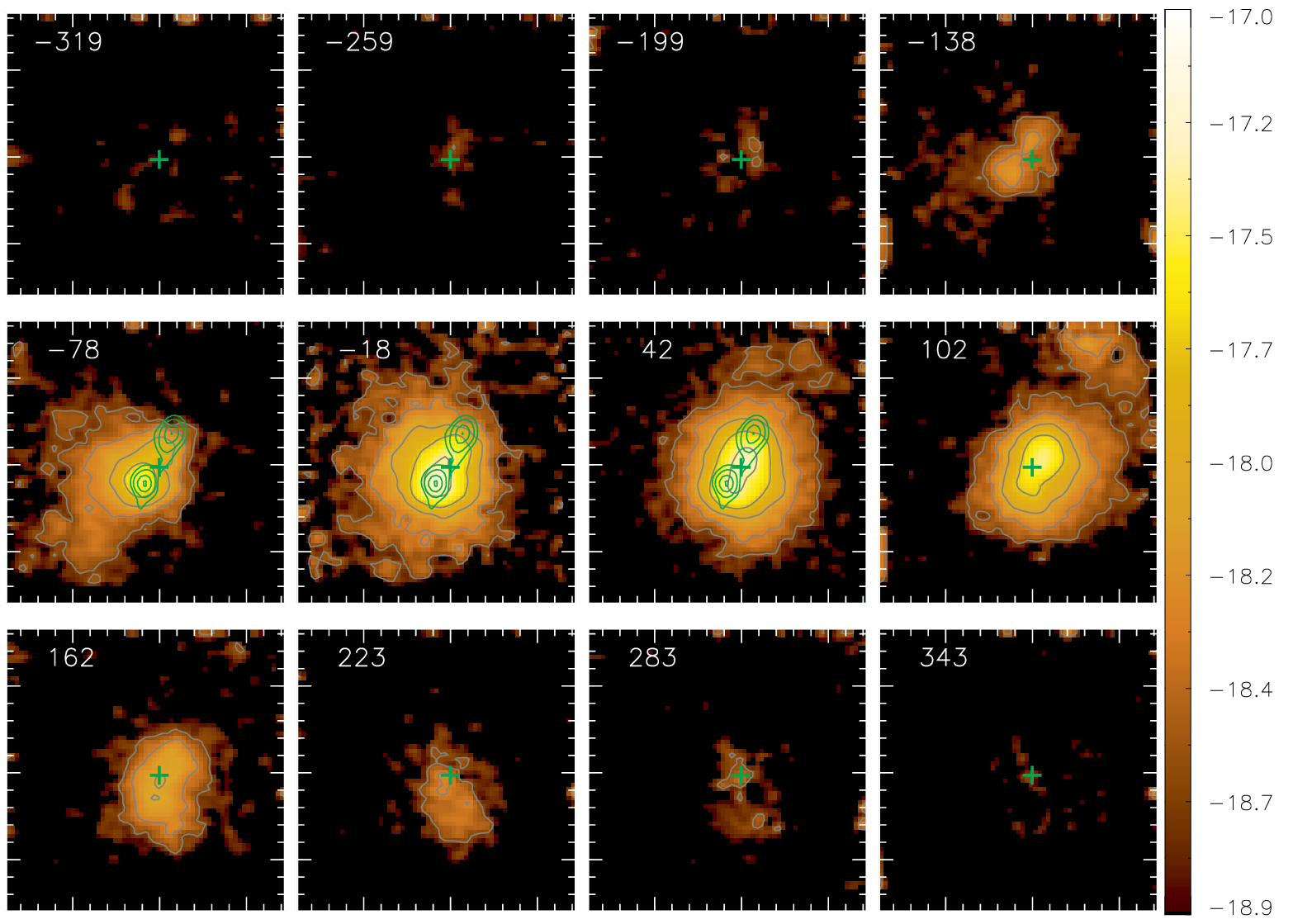} 
  \caption{Same as Fig.~\ref{slice_fe} for the  \h2\ emission line for a velocity bin of 60~\kms.} 
 \label{slice_h2}  
 \end{figure*}


Channel maps along the emission line profiles are shown in Figures~\ref{slice_fe}, \ref{slice_pb} and  \ref{slice_h2}
for the \feii, \pb, and \h2\ emitting gas. Each panel presents the flux distribution in logarithmic units integrated within the velocity bin centred at the velocity shown in the top-left corner (relative to the systemic velocity of the galaxy). The central cross marks the position of the nucleus and the green contours overlaid to some panels are from the 3.6~cm radio image of \citet{nagar99}. We do not show channel maps for \pii\  because they are similar and noisier than the \feii\ channel maps. We also do not show channel maps for the coronal lines because they do not bring additional information to that already contained in the velocity fields shown in Figs. \ref{vel} and \ref{sig}.

In Figure~\ref{slice_fe}, the channel maps along the \feii\ emission-line profile show the flux distributions integrated within velocity bins of 76~\kms (corresponding to three spectral pixels). The highest blueshifts, which reach $-$400~\kms, are observed  at 0\farcs4 north-west of the nucleus, approximately coincident with the radio hot spot, while the highest redshifts with similar velocities (400~\kms) are  observed to both sides of the nucleus (at 0\farcs4 north-west and 0\farcs6 south-east) along the orientation of the radio structure. A good correlation between the radio image and the \feii\ flux distribution is observed at velocities close to zero, where the \feii\ emission shows two peaks, at  0\farcs4 north-west and 0\farcs2 south-east of the nucleus, at the location of the two radio hot spots.
  
Figure~\ref{slice_pb} shows the channel maps for the \pb\ emitting gas for the same velocity bin as for \feii. The highest blueshifts and redshifts, reaching $\approx$300\,\kms,  are observed to both sides of the nucleus along the orientation of the radio jet (to north-west and south-east of the nucleus). The correlation between the radio and line emission observed for the \feii\ emitting gas for low velocities is not present in the \pb\ channel maps, as the emission at low velocities (between $\approx-100$ and $\approx$\,100 \kms) moves from the south-east (left and bellow the nucleus) to the north-west (right and above the nucleus), approximately following the stellar rotation pattern (blueshifts at the south-east to redshifts to the north-west of the nucleus). At the panel centred at  $-$43~\kms\ there is also a structure at $\approx$\,1\farcs6  north-east of the nucleus, still observed at the panel centred at 32~\kms\ although closer to the nucleus, and at the panel centred at 106~\kms\ there is another structure at
$\approx$\,1\farcs8 north-west of the nucleus.

The channel maps along the \h2\ emission-line profile are shown in Figure~\ref{slice_h2} for a velocity bin of 60~\kms (corresponding to two spectral pixels). These maps have some similarities to those of \pb, in the sense that at low velocities, the blueshifts are mostly observed to the east and south-east and redshifts mostly to the west and north-west, indicating rotation. But they reach lower velocities than \pb, are more uniformly distributed around the nucleus and are less correlated to the radio structure.


\section{Discussion}\label{discussion}

\subsection{Stellar kinematics}\label{disc-stel}
\begin{figure}
 \centering
 \includegraphics[scale=0.82]{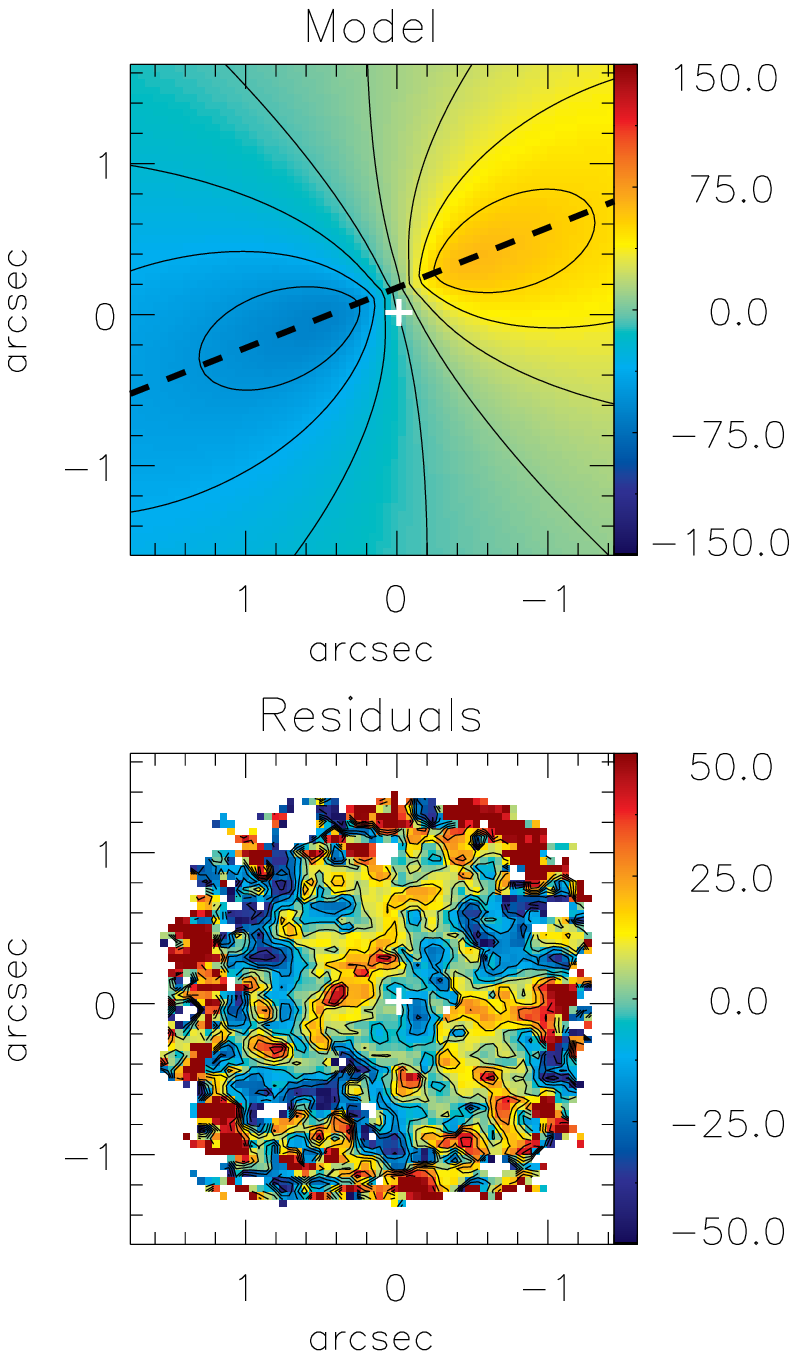} 
  \caption{Rotating disc model for the stellar velocity field (top) and residual 
map between the observed and modelled velocities. The dashed line shows the orientation 
of the line of nodes ($\Psi_0\approx112\degr$) and the cross marks the position of the nucleus. The color bar shows the range of velocities in km\,s$^{-1}$.}
 \label{plummer}  
 \end{figure}

The stellar velocity field of the central region of Mrk\,1157 (top-left panel of Fig.~\ref{stel}), suggests rotation. In order to obtain the systemic velocity,  orientation of the line of nodes, the location of the kinematical centre and an estimate for the bulge mass, we have fitted the stellar velocities with a model of circular orbits in a plane subject to a Plummer gravitational potential, in which the velocity is given by  \citep{barbosa06,n4051,mrk1066c}:

\begin{equation}
V_r=V_s + \sqrt{\frac{R^2GM}{(R^2+A^2)^{3/2}}}\frac{{\rm sin}(i){\rm cos}(\Psi-\Psi_0)}{\left({\rm cos^2}(\Psi - \Psi_0) + \frac{{\rm sin^2}(\Psi-\Psi_0)}{{\rm cos^2}(i)} \right)^{3/4}}
\end{equation}
where $R$ is the projected distance from the nucleus in the plane of the sky, $\Psi$ is the corresponding position angle, $M$ is the mass inside $R$, $G$ is the Newton's gravitational constant, $V_s$ is the systemic velocity, $i$ is the inclination of the disc ($i=0$ for a face on disc), $\Psi_0$ is the position angle of the line of nodes and $A$ is a scale length projected in the plane of the sky.

The equation above contains six free parameters [including the location of the kinematical centre $(X_0, Y_0)$], which can be determined by fitting the model to the observations. This was done using a Levenberg-Marquardt least-squares fitting algorithm, in which initial guesses are given for the free parameters.  In figure~\ref{plummer} we show the model which gave the best fit (top panel) and the corresponding residual map, obtained from the subtraction of the modelled from the observed velocities. The residuals are smaller than 30~\kms\ at most locations, indicating a good reproduction of the velocity field, but the residual map shows some systematics. In particular, the ``blueshifted strip" crossing the middle of the residual map and the  partial ``redshifted strip" to the left of it, resemble the residuals we have observed in the stellar velocity field of Mrk\,1066, which we have attributed to the presence of an oval distortion or bar \citep{mrk1066c}. This may also be the case of Mrk\,1157, since it seems to have also a nuclear bar, as found by  \citet{malkan98} using HST broad band images. The stellar velocity field (Fig.\,\ref{stel}) also suggests the presence of a small  S-distortion in the ``zero-velocity curve'', a signature of the presence of an asymmetry in the gravitational potential \citep{combes95,emsellem06,mrk1066c}.

The parameters derived from the fit are: the systemic velocity corrected to the heliocentric reference frame 
$V_s=$4473\,$\pm$\,8\,km\,s$^{-1}$, $\Psi_0=$112$^\circ$\,$\pm$\,5$^\circ$, $M=$5.4$\pm$0.5\,$\times$\,10$^8$\,M$_\odot$, $i=46\degr\pm6^\circ$, and $A=$119\,$\pm$\,18\,pc.  The derived kinematical centre is $X_0=-$1.5$\pm$13\,pc and $Y_0$=69$\pm$13\,pc measured relative to the position of the peak of the continuum emission. The systemic velocity is about 25~\kms\ smaller than the one quoted in \citet{veron01} and 75~\kms\ smaller then that of \citet{springob05}. We attribute these differences to our smaller field of view than those of the previous works, which have used the  H\,{\sc i} 21~cm line integrated within large apertures ($>$3$^\prime$) to derive $V_s$.  The inclination of the disc is in good agreement with the one obtained via the expression valid for a thin disk: $i={\rm acos}(b/a)\approx44\degr$, where $a=1.3^\prime$ and $b=1^\prime$ are the major and minor axes of the galaxy, respectively, as quoted in NASA/IPAC Extragalactic Database\footnote{http://nedwww.ipac.caltech.edu}.
The scale length $A$ and the bulge mass $M$  are similar to those obtained for the central region of other Seyfert galaxies using similar modelling \citep[e.g.][]{barbosa06,mrk1066c}.

The stellar velocity dispersion map (top-right panel of Fig.\,\ref{stel}) shows a partial ring of low $\sigma_*$ values ($\sigma_*\approx50-60\,$\kms) at $\approx$\,0\farcs6--\farcs8 from the nucleus ($\approx$\,230\,pc) immersed in a background of $\sigma_*\approx100\,$\kms\ of the bulge stars.  The presence of the partial ring is evidenced by an increase in $\sigma_*$ of about 40\,\kms, from locations at 0\farcs6  to 1$^{\prime\prime}$ from the nucleus, while the largest $\sigma_*$ uncertainties are smaller than 20\,\kms.   Similar low-$\sigma_*$ structures have been observed around the nuclei of other Seyfert galaxies and attributed to colder regions with more recent star formation than the underlying bulge \citep{barbosa06,n4051,n7582,mrk1066c}. In particular, a recent study of the stellar population in the central region of  Mrk\,1066 (which shows a similar low-$\sigma_*$ ring to that  of Mrk\,1157)  confirms the above interpretation, since the stellar population -- as derived from spectral synthesis --  in the low-$\sigma_*$ ring  of Mrk\,1066 is dominated by intermediate age (10$^8$ - 10$^9$ yr) stars \citep{mrk1066b}. We have also performed stellar population synthesis for Mrk\,1157, which indicates a similar result to that obtained for Mrk1066:  the stellar population in the low-$\sigma_*$ regions are also dominated by intermediate age stars \citep{rogerio11}.

The bulge stellar velocity dispersion can be used to estimate the mass of the super-massive black hole in the centre of Mrk\,1157:
\begin{equation}
 {\rm log}(M_{BH}/{\rm M_\odot})=\alpha+\beta\,{\rm log}(\sigma_*/\sigma_0),
\end{equation}
where $\alpha=8.13\pm0.06$, $\beta=4.02\pm0.32$  and $\sigma_0=200$\,km\,s$^{-1}$ \citep{tremaine02}. We adopt $\sigma_*\approx100\pm10$\,km\,s$^{-1}$ as representative of the bulge, a value in good agreement with previous optical measurements \citep{nelson95}, and obtain $M_{BH}=8.3^{+3.2}_{-2.2}\times10^6$\,M$_\odot$.

\subsection{Gaseous Excitation}\label{disc-excitation}

\begin{figure}
\centering
\includegraphics[scale=0.85]{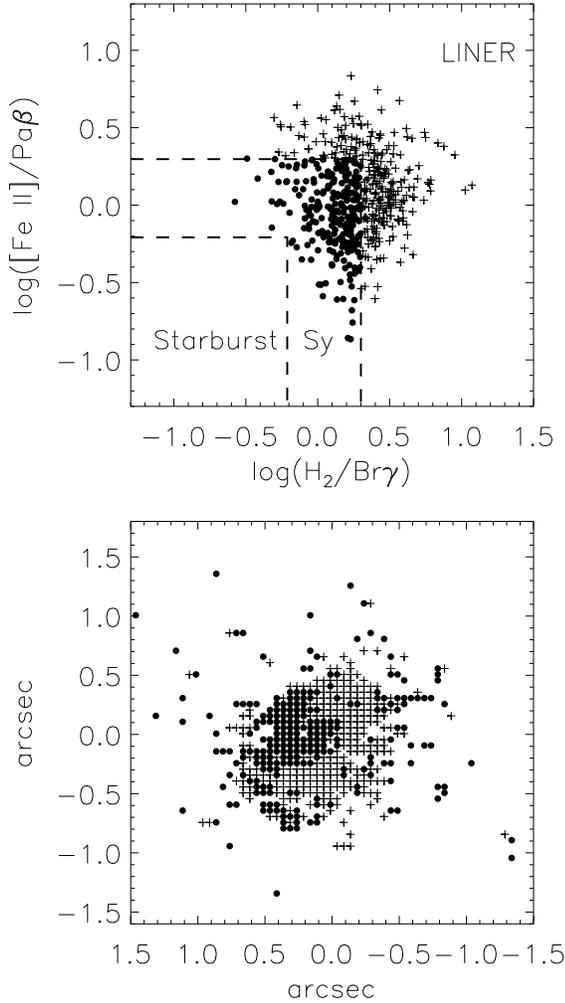}
\caption{Top: \feii$\lambda1.25\mu$m/\pb~$vs$~\h2$\lambda2.12\mu$m/\br~ line-ratio diagnostic diagram. The dashed lines delimit regions with ratios typical of Starbursts, Seyferts (filled circles) and LINERs (crosses). Bottom: Spatial distribution of the emission-line ratios of the top diagram.}
\label{diagn}
\end{figure}

In order to investigate the gaseous excitation, we constructed a spectral-diagnostic diagram \feii$\lambda1.25\mu$m/\pb~$vs$~\h2$\lambda2.12\mu$m/\br, originally proposed for integrated spectra \citep{larkin98,ardila04,ardila05}. This diagram can be used to distinguish ratios of Seyferts, LINERS and Starbursts. In particular, Seyfert nuclei have values in the range 0.6$\lesssim$\feii/\pb$\lesssim$2.0 and 0.6$\lesssim$\h2/\br$\lesssim$2.0 \citep[e.g.][]{ardila05}. Larger values are observed in LINERs and smaller values in Starbursts and H\,{\sc ii} regions. The diagram is shown in the top panel of Fig.\,\ref{diagn}, in which the Seyfert-like values are represented by filled circles and LINER-like values by crosses. In the bottom panel of Fig.\,\ref{diagn}, we show how these values are distributed spatially in the galaxy. The nucleus has a Seyfert-like value, as well as the emitting gas to the east, in regions away from the radio jet. In regions co-spatial with the radio jet, the line ratios are LINER-like, and can be interpreted as being due to an additional emission of the \h2\ and \feii\ relative to the hydrogen recombination lines due to shocks between the radio jet and the gas. This conclusion is supported by the \feii/\pb\ and \h2/\br\  line-ratio maps, as well as by the  \pii/\feii\ map (Fig.\,\ref{ratio}), which show the highest values in regions co-spatial with the radio structure (see discussion bellow).

\subsubsection{The H$_2$ emission}

The molecular hydrogen excitation can be investigated using its flux distribution, line ratios and kinematics and has been subject of several recent studies \citep[e.g.][] {sb99,reunanen02,ardila04,ardila05,davies05,riffel06,n4051,n7582,mrk1066a,sanchez09,ramos-almeida09,hicks09,sb09,sb10,guillard10}. In summary, these studies support that the the H$_2$ emission lines can be excited by two mechanisms: (i) fluorescent excitation through absorption of soft-UV photons (912--1108 \AA) in the Lyman and Werner bands \citep{black87} and (ii) collisional excitation due to heating of the gas by shocks, due to interaction of a radio jet with the interstellar medium \citep{hollenbach89} or by X-rays from the central AGN  \citep{maloney96}. The second mechanism is usually referred to  as a thermal process since it  involves the local heating of the emitting gas, while the first one is usually called a non-thermal process.

We can use the \h2$\lambda$2.2477/$\lambda$2.1218 ratio to distinguish between thermal and non-thermal processes for the H$_2$ emission of Mrk\,1157. For fluorescent excitation, typical values for this ratio are $\sim0.55$, while for thermal processes typical values are  $\sim0.1-0.2$ \citep[e.g.][]{mouri94,reunanen02,ardila04,sb09}.  As observed in Tab.\,\ref{fluxes}, \h2$\lambda$2.2477/$\lambda$2.1218$=0.11\pm0.05$ for the nucleus and  \h2$\lambda$2.2477/$\lambda$2.1218$=0.12\pm0.03$ at 0\farcs4 north-west from it. Similar values are observed for all locations of the observed field. This result indicates that the  \h2\ emission is due to thermal processes, as observed for other Seyfert galaxies \citep[e.g.][]{ardila04,ardila05,riffel06,n4051,mrk1066a,sb09}.

\begin{figure}
\centering
\includegraphics[scale=0.45]{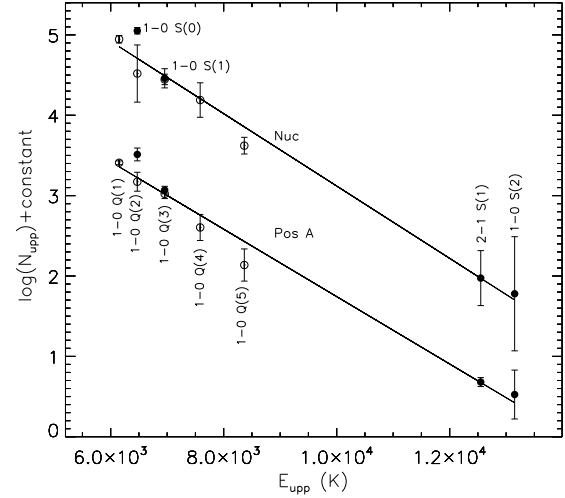}
\caption{Relation between $N_{\rm upp} = \frac{F_i \lambda_i}{A_i g_i}$ and $E_{\rm upp}=T_i$ for the H$_2$ 
emission lines for thermal excitation at the nuclear position and at 0\farcs4\,north-west 
from the nucleus. {\it Ortho} transitions are shown as filled circles (identified in the top fit) and {\it para} transitions as  open circles  (identified in the bottom fit).
}
\label{h2_temp}
\end{figure}

The thermal excitation temperature of the \h2\ emission can be obtained from the fluxes of its K-band emission lines, through the following relation \citep[e.g.][]{wilman05,sb09,mrk1066a}:

\begin{equation}
 {\rm log}\left(\frac{F_i \lambda_i}{A_i g_i}\right)={\rm constant}-\frac{T_i}{T_{\rm exc}},
\end{equation}
where $F_i$ is the flux of the $i^{th}$ H$_2$ line, $\lambda_i$ is its wavelength, $A_i$ is the spontaneous emission coefficient, $g_i$ is the statistical weight of the upper level of the transition, $T_i$ is the energy of the level expressed as a temperature and $T_{\rm exc}$ is the excitation temperature. This relation is valid for thermal excitation, under the assumption of an {\it ortho:para} abundance ratio of 3:1. In Fig.\,\ref{h2_temp} we present the observed values for $N_{\rm upp}=\frac{F_i \lambda_i}{A_i g_i}$ (plus an arbitrary constant) $vs$ $E_{\rm upp}={T_i}$ for the nuclear spectrum and for the spectrum at position A in Fig.\,\ref{large}  (0\farcs4~north-west from the nucleus). The fit of the above relation is shown in Fig.\,\ref{h2_temp} as lines and resulted in an excitation temperature of  $T_{\rm exc}=2220\pm50$\,K for the nucleus and $T_{\rm exc}=2380\pm45$\,K  for the position A. As the observed fluxes are well reproduced by the fit of the above equation, it can be concluded that the H$_2$ emitting gas is in thermal equilibrium at $T_{\rm exc}$ and that the {\it ortho} to {\it para} ratio is indeed 3, as assumed.
 
The result above thus supports thermal excitation for \h2. The heating can be provided by nuclear X-rays and/or by shocks due to the interaction of the radio jet with the interstellar medium. How can we distinguish between these two mechanisms? Using the spatial information and its correlation with the radio: the highest intensity levels of the H$_2$ emission are elongated following the orientation of the radio structure (bottom-left panel of Fig.\,\ref{flux}), what is also observed in some of the H$_2$ channel maps of Fig.~\ref{slice_h2}. Additionally, the diagnostic diagram (Fig.\,\ref{diagn}) and the \h2/\br\ line-ratio map (bottom-right panel of Fig.\,\ref{ratio}), show that the  highest values for \h2/\br\ of up to $\approx7$, are seen in regions which are co-spatial with the radio emission, suggesting that the radio jet plays a role in the H$_2$ excitation. 

The role of radio jets and X-rays in the $H_2$ emission of active galaxies was investigated  by \citet{quillen99}  using HST H$_2$ images of a sample of 10 Seyfert galaxies, as well as their radio 6~cm and hard X-ray fluxes. They found no correlation with X-rays  and  a weak correlation with radio 6-cm, suggesting that no single mechanism is likely to be responsible for the \h2\ excitation in Seyfert galaxies. This seems to be the case also for Mrk\,1157. Although we found some spatial correlation between the \h2\ flux and line ratios involving the \h2\ lines and the radio map,  we cannot exclude X-ray heating because the H$_2$ emission is less spatially correlated with the radio jet than both the \feii\ and \pb\ emitting gas. It has also smaller  $\sigma$ values (bottom-left panel of Fig.\,\ref{sig}) than the other emission lines, being similar to the stellar values, suggesting that the H$_2$ emitting gas is mainly located at the plane of the galaxy and is less affected by the radio jet than the other emission lines.

\subsubsection{The [Fe\,{\sc ii}] emission}

The excitation mechanisms of the \feii\ emission can be investigated using the \feii$\,\lambda1.2570\mu$m/\pb\ and [Fe\,{\sc ii}]$\lambda$1.2570$\,\mu$m/[P\,{\sc ii}]$\lambda$1.8861$\,\mu$m line-ratio maps shown in Fig.\,\ref{ratio}.  The first ratio is  controlled by the ratio between the volumes of partially to fully ionised regions, as the \feii~emission is excited in partially ionised gas regions. In AGNs, such regions can be created by X-ray \citep[e.g.][]{simpson96} and/or shock \citep[e.g.][]{forbes93} heating of the gas.  

For Starburst galaxies,  \feii/\pb$\lesssim0.6$ and for supernovae for which  shocks are the main excitation mechanism, this ratio is larger than 2 \citep{ardila04,ardila05}. As observed in Figs.\,\ref{ratio} and \ref{diagn}, for Mrk\,1157 \feii/\pb\ ranges from 0.6 to 6, with the highest values observed at the borders of the radio structure, indicating that the excitation by radio shocks is important at  these locations. The stronger correlation between the radio emission and the \feii\ flux map  than with the other emission lines (see Fig.\,\ref{flux}), as well as the higher increase in $\sigma$ at the edge of the radio jet, support a higher contribution of  the radio jet for the excitation of  \feii\ than for  \h2. This result is in good agreement with our previous studies of the central region of Seyfert galaxies, which indicate that the radio shocks may be the main excitation mechanism of the \feii\  near-IR emission lines \citep{sb99,riffel06,mrk1066a,sb09}.

The above conclusion is also supported by the [Fe\,{\sc ii}]$\lambda$1.2570$\,\mu$m/[P\,{\sc ii}]$\lambda$1.8861$\,\mu$m line-ratio map (bottom-left panel of Fig.\,\ref{ratio}).  These two lines have similar excitation  temperatures, and their parent ions have similar ionisation potentials and radiative recombination coefficients. Values larger than 2  indicate that shocks have passed through the gas destroying the dust grains, releasing the Fe and thus enhancing its abundance and thus emission \citep{oliva01,sb09,mrk1066a}. For supernova remnants, where shocks are the dominant excitation mechanism [Fe\,{\sc ii}]/[P\,{\sc ii}] is typically  higher than 20 \citep{oliva01}. In locations near the tips of the radio jet, Mrk\,1157 presents \feii/\pii\ values of up to 10, indicating that shocks  are important at these locations in agreement with the highest values obtained for the \feii/\pb\ at the same locations. In other regions, typical values are [Fe\,{\sc ii}]/[P\,{\sc ii}]$\approx4$, indicating a lower contribution from the radio jet and higher contribution from X-ray heating.

\subsubsection{The coronal-line emission}

Coronal lines are forbidden transitions from highly ionised species whose emission extends from the unresolved nucleus up to distances between a few tens to a few hundreds of parsecs, with the emission-line profiles usually presenting blue wings and being broader than low ionisation lines \citep[e.g.][]{ardila06,sb09,mazzalay10}. 

In the case of Mrk\,1157,  the coronal-line emission of [Ca\,{\sc viii}]\,$\lambda2.3220\,\mu$m and [Si\,{\sc vii}]\,$\lambda2.4833\,\mu$m are spatially resolved by NIFS observations, as their flux distributions (bottom-middle and bottom-right panels of Fig.~\ref{flux}) present spatial profiles with FWHM$\approx$0\farcs4, which is about 3 times larger than that of the standard star. Although the extent of the coronal-line emission region of Mrk\,1157 is much smaller than the extent of the low-ionisation region, its flux distributions are more extended along the radio jet, following the same orientation of the low-ionisation region, suggesting the same origin. This result supports 
an origin for the coronal lines in the inner part of the NLR, as suggested by previous authors \citep{ardila06}. This interpretation is also supported by the kinematics of the coronal gas, which present similar centroid velocities (Fig.\,\ref{vel}) and velocity dispersions (Fig.\,\ref{sig}) to that of the low-ionisation lines. In addition, the $h_3$ maps for both coronal lines show high negative values of up to $-0.4$, indicating the presence of blue wings, in good agreement with previous studies of other Seyfert galaxies \citep[e.g.][]{ardila06}.

\subsection{Gaseous Kinematics}\label{disc-gas}

\begin{figure*}
 \centering
 \includegraphics[scale=0.8]{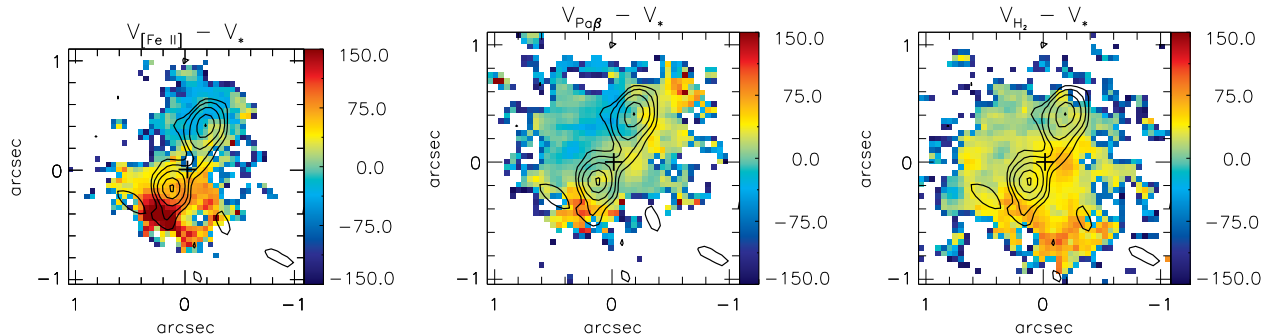} 
 \caption{Residual maps between the gaseous and stellar velocity fields.  The crosses mark the position of the nucleus, green contours are from the 3.6~cm radio image of \citet{nagar99} and the color bars show the range of residual velocities in km\,s$^{-1}$.} 
 \label{res}  
 \end{figure*}
 
As observed in Fig.\,\ref{vel} the gaseous velocity field presents blueshifts to the east and redshifts to the west, suggestive of rotation, although with the orientation of the rotation axis somewhat tilted relative to that of the stellar rotation (top-left panel of Fig.\,\ref{stel}). There are also differences between the kinematics of the molecular and ionised gas, and between that of the ionised H and \feii\ emitting gas. The difference relative to the stellar kinematics is largest for \feii, as the orientation of the apparent rotation axis differs from the stellar one by more than 45$^\circ$. 

The gas kinematics and the differences in the velocity fields of the different gas species can be better understood by inspection of the channel maps of Figs.~\ref{slice_fe}, \ref{slice_pb} and \ref{slice_h2}. Fig.\,\ref{slice_fe} shows that the \feii\ emission presents the highest blueshifts to the north-west and not to south-east, as expected for the rotation pattern followed by the stars (Fig.\,\ref{stel}). This indicates that these highest velocities are not due to rotation. The highest blueshifts, of up to $-400$\,\kms\ are observed at the location of the 0\farcs4 NW radio hot spot. At lower blueshifts, there is \feii\ emission also to the south-east, which could thus be attributed to rotation in the galaxy plane. At velocities close to zero the emission is somewhat more uniformly distributed around the nucleus, although it seems to still keep some spatial correlation with the radio map. At low redshifts the most extended emission drifts to the west, consistent with the rotation contribution, while the remaining redshifts to the south-east are observed at the location of the SE hot spot, although being spread also to the west of this hot spot. At the highest redshifts, there is still emission from the structures to the south and from the region of the hot spot at the north-west. 

The \feii\ kinematics can be understood as due to the combination of the contribution from gas in rotation in the galaxy plane and outflows associated with the radio jet, which has the north-west part somewhat tilted towards us (in order to produce blueshifts to the north-west). At low velocities, the \feii\ kinematics has an important contribution from rotation, as evidenced by the emission structure moving from south-east  at negative velocities to north-west of the nucleus at positive velocities, approximately following the orientation of the line of nodes of the stellar velocity field. At high velocities the outflows dominate, pushed by the radio jet. Nevertheless, the spatial correlation with the radio jet seems to be present also in the lower velocity channels, suggesting that there is contribution from the radio jet in driving lower velocity kinematics as well. This is also supported by the high velocity dispersion observed over the whole region covered by the \feii\ emission (Fig.\,\ref{sig}).

 What is the orientation of the outflow relative to the galaxy plane? Inspection of the HST image of Mrk\,1157 in combination with the observed kinematics, indicates that, if the spiral arms are trailing, the near side of the galaxy is the north--north-east (line of nodes at PA=112$^\circ$) and the far side is the south--south-west. The near (blueshifted) side of the outflow can thus have three orientations relative to the plane: (i) be along the plane, (ii) in front of it or (iii) behind it. If it is not along the plane, in order to appear extended in the sky, and to show blueshifts to the north-west, it should make an angle smaller than 45$^\circ$ with the plane, as this is the inclination of the galaxy plane (relative to the plane of the sky), as obtained from the fitting of the stellar velocity field. Inspection of the \feii\ emission distribution at the highest redshifts ($350-426$\,\kms in the channel maps) reveals also some emission to the north-west at the location of the NW hot spot. This suggests that the outflows (and the radio jet), although having the north-west side somewhat tilted towards us, is launched close to the plane of the sky, so that, if the outflow opens in a cone, the blueshifts would come from the front wall of the cone and the redshifts from its back wall. If this is the case, the cone axis would be behind the plane to the north-west, with part of the  wall of the cone in front of the plane depending on the launching angle and the cone opening angle. In Fig.\,\ref{scenario} we present a sketch diagram for the geometry of the outflow relative to the plane of the galaxy and plane of  the sky. To the south-east, the cone axis outflow would be in front of the plane. One difficulty with this scenario is the lack of emission to the south-east at the highest blueshifts, but this could be due to an asymmetry in the gas distribution along the path of the outflow. The hypothesis that the radio jet and associated outflow is oriented close to the plane of the sky is supported by the \pb\ channel maps (Fig.\,\ref{slice_pb}), which shows, at the highest blueshifts and redshifts, extended emission to both sides of the nucleus  following the orientation of the radio jet. At lower velocities, the main contribution is from gas rotating in the plane, from blueshifts to the south-east to redshifts to the north-west.

The \h2\ channel maps (Fig.\,\ref{slice_h2}) reach smaller blueshifts but are similar to those of \pb\ for velocities between $\approx-100$\,\kms\ and 100\,\kms, consistent with rotation in the galaxy plane. At the highest redshifts there is some emission to the south--south-west, at the same location where there is some emission also from \feii\ at similar redshifts. There is less correlation with the radio structure than for \pb.

Distinct kinematics between the \h2\ and \feii\ kinematics seem to be a common characteristic of active galaxies, having been observed in many other Seyfert galaxies  \citep{sb99,ardila04,ardila05,riffel06,n4051,n7582,sb10,mrk1066c}.



\subsubsection{Comparison between the stellar and gaseous kinematics}

In order to map differences between the stellar and gaseous kinematics, we have constructed the residual maps of Fig.\,\ref{res}, obtained from the difference between the centroid \feii, \pb\ and \h2\ velocities and the stellar velocity field. 

The largest and clearest differences are observed in the \feii\ residuals, which show redshifts of up to 130\,\kms\ to the south-east and blueshifts of up to $-$80\,\kms\ to the north-west of the nucleus in clear association with the radio jet. 

The \pb\ residuals are smaller than those of \feii\, but there are some blueshifts ($-$30\,\kms) to the north-east -- thus not associated with the radio jet --  which appear also in the channel map at $-43$\,\kms\ (Fig.\,\ref{slice_pb}). The gas excitation is higher at this location (see Fig.\,\ref{ratio}) while the velocity dispersion is lower (Fig.\,\ref{sig}). This could be a mild outflow, not shocked by the radio jet and ionised by the AGN. There are also some small redshifts (30\,\kms) to the west--south-west, which could be a counterpart of the mild blueshifted outflow and some redshifts at the tip of the SE radio hot spot, where there is also an enhancement in the gas velocity dispersion, which could be part of the the outflow to the SE pushed by the radio jet. 

The \h2\ residual maps show small redshifts mostly to the south--south-west ($\le$40\,\kms), a structure also present in the redshifted \h2\ channel maps (Fig.\,\ref{slice_h2}), which could be part of the mild outflow observed in the \pb\ kinematics, as described above.

The residual maps support the interpretation discussed above for the channel maps. The negative velocities to north-west and positive velocities to the south-east, clearly observed in the \feii\ residual map along the radio jet can be interpreted as due to the emission of outflowing gas pushed by the radio jet. 
The \feii\ emitting gas is the best tracer of interaction between ambient gas and a radio jet, as the jet destroys dust grains on its way out, releasing the Iron which produces enhanced \feii\ emission at the shock locations. The presence of shocks explains the high \feii\ velocity dispersions. Such enhancements are expected when gas is disturbed by a radio jet \citep[e.g.][]{dopita95,dopita96}.

The \feii\ residual map also supports that the north-west part of the radio jet is somewhat tilted towards us, in spite of the fact that the channel maps also suggest that the angle between the jet and the plane of the sky is small, so that redshifts can also be observed at the location of the NW hot spot.  
It can thus be concluded that the farthest \feii\ emission originates in gas which extends to high galactic latitudes. This may be the case for some \pb\ emission at the highest velocities which produces some velocity dispersion enhancement seen in Fig.\,\ref{sig}. But most of the \pb\ emission seems to originate in the galaxy plane, as revealed by the rotation pattern which dominates the emission in the \pb\ channel maps. Most of the \h2\  emission also seems to originate in gas rotating in the plane, as supported by the low velocity dispersions (Fig.\,\ref{sig}). The fact that the molecular gas is observed mostly in the galaxy plane while the ionised gas is frequently observed to higher galactic latitudes has also been observed by us for other  Seyfert galaxies  \citep[e.g.][]{riffel06,n4051,n7582,sb10,mrk1066c}. This can be understood as due to the fact that, in the plane, the molecular gas is more shielded by dust clouds from the ionizing radiation which can destroy the \h2\ molecule.

The velocity fields and $\sigma$ maps for the coronal lines  have similar values to those observed for the low ionisation species (\pii, \feii\ and \pb), supporting an origin for these lines in the inner NLR, in good agreement with the results discussed in the previous section.

\begin{figure*}
 \centering
 \includegraphics[scale=0.8]{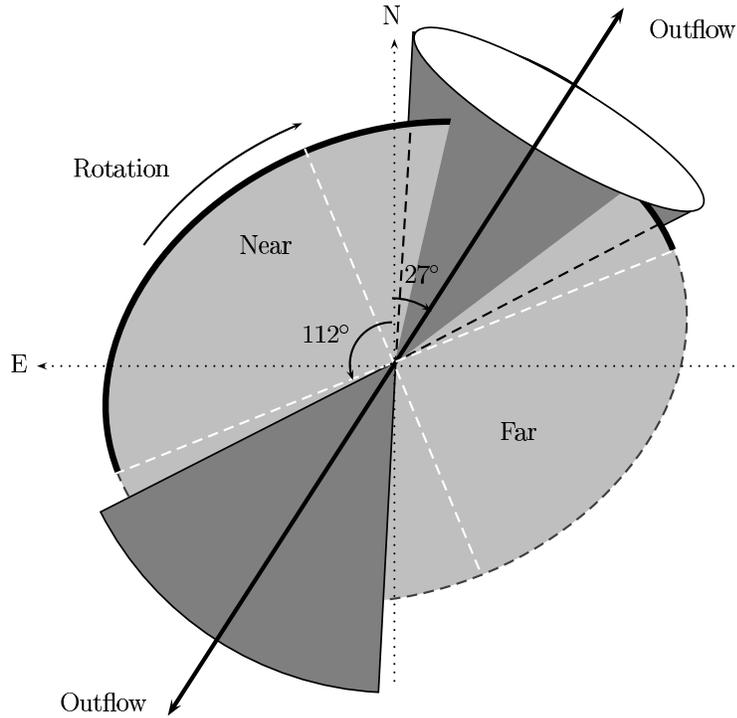} 
 \caption{Physical scenario of the inner 450\,pc of Mrk\,1157 showing the two kinematic components: rotation  in the plane of the galaxy and the outflow, represented by a bi-cone. The white-dashed lines show the orientation of the major and minor axes of the galaxy.} 
 \label{scenario}  
 \end{figure*}

In Fig.\,\ref{scenario} we present a sketch diagram for the physical scenario of the inner 450\,pc of Mrk\,1157. Two kinematic components are observed for the emitting gas. The first is due to gas located in the plane of the galaxy presenting a similar rotating pattern to that observed for the stars, with a disc inclination of $i=46^\circ$ relative to the plane of the sky and line of nodes oriented along PA$\approx112^\circ$. The second component is due to emission of the outflowing gas within a bi-cone, with opening angle of $\approx60^\circ$, oriented along PA$\approx153^\circ$, being oriented approximately perpendicular to the line of sight, making an angle of $\approx15^\circ$ with the plane of the sky.

\subsubsection{Mass outflow rate}

In order to quantify the outflows from the nucleus of Mrk\,1157, we estimate the ionised-gas mass outflow rate through a circular cross section  located at 0\farcs6 from the nucleus along a bi-cone with the same orientation of the radio jet (PA$=$153$^\circ$) and opening angle of 60$^\circ$ (estimated from the left panel of Fig.\,\ref{res}).  The component of the velocity of the outflowing gas along the bi-cone axis ($v_{\rm out}$)  is related to the observed velocity ($v_{\rm obs}$) by $v_{\rm out}=v_{\rm obs}/{\rm sin}\,\theta$, where $\theta$ is the angle between the bi-cone axis and the plane of the sky, and thus the mass outflow rate can be obtained from
\begin{equation}
 \dot{M}_{\rm out} = \frac{2 m_p N_e v_{\rm obs} f A}{sin\,\theta},
\end{equation}

\noindent where $A=\pi r^2\,\approx\,3.15\,\times10^{41}\,{\rm cm^2}$ is the area of the circular cross section, $m_p=1.67\times10^{-24}\rm{g}$ is the mass of a proton, $f$ is the filling factor, $N_e$ is the electron density and the factor 2 is included in order to account for the outflows to both sides of the nucleus. The filling factor can be obtained from  the relation

\begin{equation}
4\pi  L_{Pa\beta}=2.046\,\times\,10^{-26} \times N_e^2  V f
\end{equation}

\noindent where L$_{Pa\beta}$ is the \pb\  luminosity of the outflowing gas and $V$ is its volume, and the physical quantities are in $CGS$ units. We identify the outflowing gas as that corresponding to the channels at velocities higher than 200\,\kms ($V\ge$200\,\kms and $V\le\,-$200\,\kms), adding the fluxes of the corresponding channel maps plus half of the fluxes in the maps centred at $-191$\,\kms\ and 181\,\kms\, (the other 50\% being attributed to emission from gas in the plane of the galaxy), which is $1.3\,\times\,10^{-15}$\,erg\,s$^{-1}$\,cm$^{-2}$, resulting in $L_{Pa\beta} = 5.8\times10^{38}$\,erg\,s$^{-1}$. The volume of the outflowing gas was estimated as a cone with height  $h=0.6$\arcsec = 177.6\,pc and opening angle of 60$^\circ$, resulting in $V=5.75\,\times\,10^{61}{\rm \,cm^{-3}}$ and $f=0.025$.

Assuming $N_e=500\,{\rm cm^{-3}}$, $f=0.025$ and  $v_{\rm out}=75$~\kms (from Fig.~\ref{res}), we obtain $\dot{M}_{\rm out}{\approx1.55/\rm sin\,\theta}~{\rm M_\odot\, yr^{-1}}$. 

As discussed above, the radio jet (and thus the bi-cone) is probably oriented close to the plane of the sky in order to account for the highest blueshifts and redshifts observed in the ionised gas emission. If $\theta=15^\circ$, $\dot{M}_{\rm out}\approx 6{\,\rm M_\odot\, yr^{-1}}$, which is of the order of the value we have recently obtained  for NGC\,4151, of  $\dot{M}_{\rm out} \approx2~{\rm M_\odot\, yr^{-1}}$ \citep{sb10}. This value is also within the range of values reported by  \citet{veilleux05} for warm ionised gas outflows from luminous active galaxies of $\dot{M}_{\rm out} \approx0.1-10~{\rm M_\odot\, yr^{-1}}$.

In \citet{barbosa09} we have reported much smaller values, in the range $\dot{M}_{\rm out} \approx(1-50)\times10^{-3}~{\rm M_\odot\, yr^{-1}}$ for a sample of six Seyfert galaxies. But in that work we have adopted an electron density 5 times lower ($N_e=100~{\rm cm^{-3}}$) and could not calculate the filling factor, adopting a value 25 times lower ($10^{-3}$). If we adopt the electron density and  $f$ values of the present work, we would increase the $\dot{M}_{\rm out} $  values by a factor of 150, and the resulting values would be in the range $\dot{M}_{\rm out} \approx\,0.13-6.25{\rm M_\odot\, yr^{-1}}$, more similar to the above. 

The same applies to the case of  \citet{n7582}, for which we used $N_e=100~{\rm cm^{-3}}$ and $f=0.01$, and the value found of $\dot{M}_{\rm out} \approx5\times10^{-2}~{\rm M_\odot\, yr^{-1}}$ should be multiplied by 12.5, resulting in $\dot{M}_{\rm out} \approx\,0.6~{\rm M_\odot\, yr^{-1}}$. In the case of \citet{mrk1066b}, the factor is just 2.5, and the corresponding revised value is $\dot{M}_{\rm out} \approx0.15~{\rm M_\odot\, yr^{-1}}$.


In the following we compare the mass outflow rate, obtained above, with the accretion rate necessary to power the AGN at the nucleus of Mrk\,1157, which can be 
obtained from

\begin{equation}
 \dot{m}=\frac{L_{\rm bol}}{c^2\eta},
\end{equation}
where $L_{\rm bol}$ is the nuclear bolometric luminosity, $\eta$ is the efficiency 
of conversion of the rest mass energy of the accreted material into radiation and $c$ 
is the light speed. The bolometric luminosity can be approximated by 
$L_{\rm bol} \approx 100L_{\rm H\alpha}$, where $L_{\rm H\alpha}$ is the \ha\ nuclear luminosity \citep[e.g.][]{ho99,ho01}. The \br\ nuclear flux measured for a 0\farcs25$\times$0\farcs25 nuclear aperture is $F_{\rm Br\gamma}\approx4.6\times10^{-16}\,{\rm erg\, s^{-1} cm^{-2}}$, as presented in Table\,\ref{fluxes}. As shown in Fig.\,\ref{ratio}, we obtain a reddening of $E(B-V)\approx1.7$ for the nucleus of Mrk\,1157 using the \br/\pb\ line ratio. Correcting the \br\ flux for this reddening using the law of \citet{cardelli89}, we obtain  $F_{\rm Br\gamma}\approx8\times10^{-16}\,{\rm erg\, s^{-1} cm^{-2}}$. Assuming a temperature $T=10^4$~K and an electron density $n_e=500~{\rm cm^{-3}}$, the ratio between  \ha\ and \br\ is predicted to be $F_{\rm H\alpha}/F_{\rm Br\gamma}=103$ \citep{osterbrock06}. Thus $L_{\rm H\alpha}\approx3.7\times10^{40}\,{\rm erg\, s^{-1}}$ for a distance to Mrk\,1157 of $D=61.1$~Mpc and the nuclear bolometric luminosity is estimated to be $L_{\rm bol}\approx3.7\times10^{42}\,{\rm erg\, s^{-1}}$. Assuming $\eta\approx0.1$, which is a typical value for a ``standard'' geometrically thin, optically thick accretion disc \citep[e.g.][]{frank02}, we obtain a mass inflow rate of $\dot{m}\approx6.5\times10^{-4}~{\rm M_\odot\, yr^{-1}}$.

The mass outflow rate ($\dot{M}_{\rm out}$) in the NLR of Mrk\,1157 is about four orders of magnitude larger than $\dot{m}$, a ratio comparable to those observed for other Seyfert galaxies (using  the  $\dot{M}_{\rm out}$ values presented above), which indicates that most of the outflowing gas observed in the NLR of active galaxies does not originate in the AGN (as the nuclear outflow rate cannot be higher than the accretion rate) but in the interstellar medium surrounding the galaxy nucleus, which is pushed away by the AGN outflow.

Finally, we can use the above mass outflow rate to estimate the 
kinetic power of the outflow. Following \citet{sb10} the kinetic power can be 
obtained from

\begin{equation}
\dot{E}\approx\frac{\dot {M}_{out}}{2}(v_{out}^2+\sigma^2),
\end{equation}
where $v_{out}=v_{obs}/sin\theta$ is the velocity of the outflowing gas
 and $\sigma$ is its velocity dispersion. Using $\sigma\approx200\,$km\,s$^{-1}$ ( from Fig.\,\ref{sig})
and $v_{out}=v_{obs}/sin\theta=75{\rm km\,s^{-1}}/sin 15^\circ=290$\,km\,s$^{-1}$ (discussed above) we obtain 
$\dot{E} \approx 2.3\times10^{41}$ erg\,s$^{-1}$. This kinetic power is in good agreement with those derived for 
the Seyfert galaxy NGC\,4151 \citep{sb10} and for the warm outflow in compact radio sources \citep{holt11,holt06,morganti05}. Comparing 
the kinetic power with the bolometric luminosity we find that $\dot{E}\approx 0.15\times L_{\rm bol}$, implying that 15\% of the 
mass accretion rate is transformed in kinetic power in the NLR outflows. This value is about one order of magnitude larger 
than the AGN feedback derived by \citet{dimatteo05} in simulations for the co-evolution of black holes and galaxies, 
in order to match the $M-\sigma$ relation and from one to two orders of magnitude higher than those derived for six Seyfert galaxies using integral field optical spectroscopy (after revising the filling factor and gas density values as discussed above) \citep{barbosa09}. Nevertheless, it should be noticed that the $\dot{E}$ derived here may have an uncertainty of up to an order of magnitude due to the uncertainty in the  geometry of the outflow.

\section{Conclusions}\label{conclusions}

We have analysed two-dimensional near-IR $J-$ and $K_l-$bands spectra from the inner $\approx$\,450\,pc radius of the Seyfert 2 galaxy Mrk\,1157 obtained with the Gemini NIFS instrument at a spatial resolution of $\approx$35~pc (0\farcs12) and velocity resolution of $\approx$\,35$-$45\,\kms. We have mapped the emission-line flux distributions and ratios, as well as the stellar and gaseous kinematics, in the molecular and ionised emitting gas. The main conclusions of this work are:

\begin{itemize}

\item The stellar velocity field is  well reproduced by a model in which the stars follow circular orbits in a Plummer gravitational potential, with a major axis at PA=112$^{\circ}\pm5^{\circ}$. The near side of the galaxy is the north--north-east; 

\item The stellar velocity dispersion presents a partial ring (radius $\approx$\,250\,pc) of low values (50$-$60\,\kms) surrounded by higher $\sigma_*$ values from the bulge stars. This low-$\sigma_*$ ring is interpreted as being originated in colder regions with more recent star formation which still keep the lower velocity dispersion of the gas from which the stars have formed; 

\item The stellar velocity dispersion of the bulge ($\approx$\,100\,\kms) implies a black hole mass of $M_{BH}=8.3^{+3.2}_{-2.2}\times10^6$\,M$_\odot$;

\item Emission-line flux distributions of molecular hydrogen and low-ionisation gas are most extended along PA$=27/153^\circ$ in agreement with previous optical \oiii\ imaging and following the radio jet. This emission extends to at least $\approx$\,450\,pc (1\farcs5) from the nucleus. The coronal lines of [Ca\,{\sc viii}] and [Si\,{\sc vii}] are resolved extending up to $\approx$\,150\,pc (0\farcs5) and are slightly more extended also along PA$=27/153^\circ$;

\item The reddening map obtained via the \pb/\br~line ratio presents typical values of $E(B-V)=0.5$, with the highest values of up to $E(B-V)=1.8$ seen at the nucleus;

\item  The \h2\ gas is excited by thermal processes and has an excitation temperature $T_{\rm exc}\approx2300$\,K. Its emission is mainly due to X-ray excitation from the central AGN, but a small contribution from shocks produced by the radio jet cannot  be discarded;

\item The \feii\ emission shows flux and velocity dispersion enhancements  due to shocks produced
by the radio jet which destroys the dust grains in its way out from the AGN, releasing the Fe;

\item From the comparison between the stellar and gaseous velocity fields, as well as from channel maps along the emission line profiles, we conclude that the gas has two kinematic components: (i) the first is due to gas located in the galaxy  plane, in similar rotation to that characterizing the stellar motion; this component dominates the \h2\ emission; (ii) the second component is in outflow, which is oriented along the radio jet at PA=153$^\circ$ with the near (blueshifted) side to the north-west, but is launched close to the plane of the sky, making an angle less than $\approx$\,30$^\circ$ with the galaxy plane; this component extends to high latitudes and dominates the \feii\ emission;

\item The ionised gas mass outflow rate through a cross section with radius $\approx$\,100\,pc located at a distance of $\approx$180\,pc from the nucleus is $\dot{M}_{\rm out}\approx6~{\rm M_\odot\, yr^{-1}}$;

\item Revised mass outflow rates from our previous studies of 8 other Seyfert galaxies, using the better constrained filling factors and gas densities we have determined in the present study, are in the range 0.13--6.25 M$\odot$yr$^{-1}$, thus similar to the case of Mrk\,1157. These rates are much larger than the accretion rates to the AGN, implying that most of the NLR outflow mass originates in the interstellar medium of the galaxy, which is pushed by a less massive nuclear outflow.  

\item The coronal gas, although being less extended, has similar kinematics to that of the low-ionisation gas, suggesting an origin in the inner NLR.

\end{itemize}
 
Our final statement, which applies to all Seyfert galaxies we have recently  observed with NIFS, is that the \h2\ and the \feii\ emitting gas sample different flux distributions and kinematics, indicating a distinct origin. The molecular gas is in rotation in the plane of the galaxy, where is the source of fuel for  the super-massive black hole. The \feii\ emission, on the other hand, has higher velocity dispersions and extends to high latitudes, mapping mostly the AGN outflows.

\section*{Acknowledgements}
 We thank the referee, Joanna Holt, for valuable suggestions which helped to improve the present paper.
This work is based on observations obtained at the Gemini Observatory, 
which is operated by the Association of Universities for Research in Astronomy, Inc., under a cooperative agreement with the 
NSF on behalf of the Gemini partnership: the National Science Foundation (United States), the Science and Technology 
Facilities Council (United Kingdom), the National Research Council (Canada), CONICYT (Chile), the Australian Research 
Council (Australia), Minist\'erio da Ci\^encia e Tecnologia (Brazil) and south-eastCYT (Argentina).  
This research has made use of the NASA/IPAC Extragalactic Database (NED) which is operated by the Jet
 Propulsion Laboratory, California Institute of  Technology, under contract with the National Aeronautics and Space Administration.
This work has been partially supported by the Brazilian institution CNPq.

\label{lastpage}

\end{document}